\documentclass[12pt]{article}
\usepackage[a4paper, margin=0.8in]{geometry}
\usepackage{mathrsfs}
\usepackage{color}
\usepackage{psfrag}
\usepackage{graphicx,epstopdf}
\usepackage{caption,subcaption}
\usepackage{enumerate}
\usepackage{amsmath,amssymb,calc,amsfonts, mathrsfs}
\usepackage{latexsym}
\def\beq{\begin{eqnarray}}
\def\eeq{\end{eqnarray}}

\def\={\stackrel{\Delta}{=}}

\def\lie{\pounds}
\def\mcD{\mathcal{D}}
\def\mcS{\mathcal{S}}
\def\mcR{\mathcal{R}}

\DeclareMathOperator\erf{Erf}
\usepackage[utf8]{inputenc}
\usepackage{graphicx}
\usepackage{authblk}
\usepackage{scalerel}
\title{Spherical trapped surfaces in $n$-dimensional general relativity}
\author[1]{Ayan Chatterjee\footnote{ayan.theory@gmail.com}
} 
\author[2,3]{Suresh C. Jaryal\footnote{suresh.fifthd@gmail.com}}
\author[1]{ Akshay Kumar\footnote{akshay.relativity@gmail.com}
}
\affil[1]{Department of Physics and Astronomical Science, Central University of Himachal Pradesh, Dharamshala- 176215, India. }
\affil[2]{Department of Physics and Photonics Science,
National Institute of Technology Hamirpur, Hamirpur, Himachal Pradesh- 177005, India.}
\affil[3]{Department of Physics and Astronomical Sciences,
Central University of Jammu,
Samba, Jammu and Kashmir- 181143, India.}
\begin{document}
\date{}
\maketitle
\begin{abstract}
In this paper, we examine gravitational collapse of matter fields in $n$-dimensional
general relativity. The matter energy-momentum tensor under consideration includes dust, 
perfect fluids with equations of state and matter admitting
bulk and shear viscosity. By adjusting various parameters of the matter energy-momentum tensor,
we determine the trapped region and 
spherical marginally trapped surfaces in homogeneous and inhomogeneous models of collapse.
We show that, as expected, the time development of marginally trapped tube
is intricately related to the initial velocity and density profiles of the collapsing matter configuration.
This study clarifies the role of initial data in the formation 
of spacetime singularity during gravitational collapse and
implies that, under generic conditions on 
the matter profiles, the central spacetime singularity is always covered by a horizon. 
\end{abstract}
%%%%%%%%%%%%%%%%%%%%%%%%%%%%%%%%%%%%%%
%%%%%%%%%%%%%%%%%%%%%%%%%%%%%%%%%%%%%%%%%%%%%%%%%%%%%%%%%%%%%%%%%%%%%%%%%%
\section{INTRODUCTION}\label{sec1}
The foundational aspects of formation of black hole horizon, its 
relation to trapped regions and spacetime singularities, remain an open problem \cite{Hawking_Ellis, Wald,joshi,Joshi:2012mk, Clarke}.  
A theorem by Penrose shows that
under certain energy conditions, if trapped surfaces form during gravitational collapse, then a
spacetime singularity is inevitable in general relativity \cite{Penrose:1964wq, Penrose:1969pc}. 
Indeed, matter of sufficiently high mass eventually overcomes quantum degeneracy pressure leading to the formation of a black hole, indicating that singularities are a natural consequence of our classical theory
of gravity \cite{Landau_Lifshitz}.  It is also hypothesized that cosmic censorship \cite{Penrose:1969pc} 
necessarily implies that a future horizon must exist to cover the central singularity. 
This hypothesis is among the unsolved problems of classical gravitational physics. 
It is believed that the study of a wider class of gravitational collapse models may shed light into 
the precise formulation of the problem and steer toward an approach to its
 solution \cite{Wald,joshi,Joshi:2012mk, Clarke, Penrose:1969pc}. In this regard, the pioneering work of  Oppenheimer-Snyder, Dutt, Tolman, Bondi, Misner-Sharp-Hernandez \cite{SD,OS,Lemaitre:1933gd, Tolman:1934za, Bondi:1947fta, Misner_Sharp, Hernandez:1966zia} are also useful. However, the time development of the trapped region itself and its boundary remains to be understood under general conditions. In this paper,
we extend the study of gravitational collapse of generic matter configurations in higher dimensions, emphasizing the particular role initial velocity and initial density profiles have, in governing the genesis 
and development of spacetime singularities and the horizon.

The notion of trapped surfaces is fundamental in this study, and several proposals 
have emerged that have proved useful \cite{Hayward:1993wb, 
Ashtekar:2004cn,Booth:2005qc}. In this list, the notion of marginally 
trapped tube (MTT) is natural in addressing a 
broad spectrum of topics related to the behavior of horizons  \cite{Ashtekar:2005ez}. 
The definition is as follows:
In an $n$-dimensional spacetime, MTT is a $(n-1)$-dimensional 
hypersurface foliated by closed $(n-2)$-dimensional spacelike
spheres, such that the expansion scalar of the outgoing null vector field
($\ell^{\mu}$)  is zero $\theta_{(\ell)}=0$, and the ingoing 
null vector field ($n^{\mu}$) has negative expansion $\theta_{(n)}<0$.
Note that this definition of horizon is independent of the signature of MTT, and hence, 
naturally reproduces the entire span of horizon development. The tangent vector to 
the MTT, given by $t^{\mu}=(\ell^{\mu}-C\, n^{\mu})$, may be spacelike, timelike or null, 
depending on the sign of $C$. Spacelike MTTs accurately describe a dynamically 
evolving horizon (in which case matter crosses the horizon), whereas a null MTT faithfully 
reproduces the behavior of 
an isolated, stationary horizon \cite{Dreyer:2002mx,Schnetter:2006yt, Booth:2005ng,Chatterjee:2020khj,Chatterjee:2021zre,Jaryal:2022rzd}.
This improved description of horizon has several interesting 
ramifications for the classical and quantum theory of black hole horizons. First, it 
has led to a proof of the laws of black hole mechanics  
on the covariant phase space for null isolated and conformal Killing horizons \cite{Ashtekar:2000sz,Ashtekar:2000hw, Chatterjee:2008if,Chatterjee:2014jda,Chatterjee:2015fsa}. Second,
the time development of dynamical horizons and the horizon flux formula arises naturally out of this
geometry of MTT \cite{Ashtekar:2002ag, Ashtekar:2003hk, Chatterjee:2012um}. Thirdly, in 
the numerical study of mergers of compact objects, 
the formulation of MTT is used to track 
the horizon formation \cite{Ashtekar:2004cn,Dreyer:2002mx,Schnetter:2006yt}. Fourth, 
the quantum mechanical understanding of black hole entropy in loop quantum gravity is crucially 
dependent on the quantum properties of fields on MTT \cite{Ashtekar:1997yu, Ashtekar:2004eh,Kaul:2000kf,Chatterjee:2020iuf,Perez:2017cmj}.  
The marginal trapped surfaces (MTS) foliating the MTTs are also useful in developing a concrete
understanding of gravitational collapse, including horizon formation. 
If the horizon is isolated or dynamically evolving, the MTS is ideally suited to portray the
instantaneous quasilocal properties of such horizons \cite{Booth:2005ng,Chatterjee:2008if,Chatterjee:2014jda,Chatterjee:2015fsa,krasinski_hellaby,Booth:2010eu,Bengtsson:2008jr,Bengtsson:2010tj,Bengtsson:2013hla,Booth:2012rm,Creelman:2016laj,Booth:2017fob, Schnetter:2005ea, Nielsen:2010wq, gutti2, sherif}. Indeed, in  a dynamical process of
gravitational collapse, one may track these quasilocal objects and determine the growth of horizon,
(in relation to the singularity formation) and hence, provide \emph{real time} construction of a 
black hole horizon. In the earlier studies \cite{Booth:2005ng,Chatterjee:2008if,Chatterjee:2014jda,Chatterjee:2015fsa} several matter models admitting a large variety 
of initial velocity and density profile 
have been examined. They have revealed peculiarities pertaining to the nature of these MTTs.
For example, the signature of these surfaces (denoted by $C$) has led to identification of regions 
of the parameter space for which stable horizons form. More precisely, for the spherically
symmetric gravitational collapse, the value of $C$ determines the stability of MTT under evolution.
Positive, and null values for $C$ points to stable horizons, whereas negative $C$
are unstable \cite{Andersson:2005gq,Andersson:2007fh}. It is thus possible to identify
and distinguish the mass profiles for which the collapsing matter shells contribute to stable 
horizons that are required for study of classical properties of black holes.
In short, the MTT formalism is ideally suited to address 
the gravitational collapse of matter shells leading to black holes.

In this paper, we shall use the MTT formalism to examine the intricacies of 
gravitational collapse in $n$-dimensional model in the
Einstein theory of gravity. Although gravitational collapse in this theory 
has been carried out earlier, in particular reference to the determination of naked singularities and 
the censorship conjecture \cite{joshi,Joshi:2012mk,Eardley:1978tr,Christodoulou_1984,Papapetrou,
Newman:1985gt,
 Ori:1987hg, Newman_joshi,Dwivedi:1989pt,Ori:1989ps,Lake:1991bff, Lake:1991qrk,Joshi:1993zg, Christodoulou_1994,Dwivedi:1994qs,Singh:1994tb,
Jhingan:1996jb, Harada:1998cq, Harada:1998wb, Maeda:1998hs,Joshi:1998su, Maeda:2006pm,  Goswami:2006ph, Christodoulou_1999,Dafermos:2008en}, here we focus, as the shell collapse proceeds, on the 
formation of MTTs and compare the developments of spacetime singularity with these quasilocal
horizons. Furthermore, our calculations shall encompass an extensive
set of realistic models of energy-momentum tensors including that of a viscous fluid.  As 
the collapse proceeds, and each shell of the matter configuration
falls in, we obtain (i) end state of continuous gravitational collapse, along with the time of 
formation of spacetime singularities and horizons. In particular, as the process continues, we 
track the development of the singularity and its relation to location 
of spherical marginally trapped tubes. (ii) For a broad set of initial mass profiles 
of collapsing matter, we identify the
regions of the parameter space where the MTS develops as a dynamical horizon, and compare it with its
equilibrium, isolated horizon phase.
This shall also help us to  (iii) correctly locate the boundary of the trapped region
developing during gravitational collapse. Each of these elements, (i), (ii), and (iii) is 
then repeated over other density profiles, along with a variation in 
the specification of initial velocity.
Furthermore, as pointed out previously, 
the sign of $C$ provide clues to the classical stability of MTTs, and 
hence, as we shall show, an outgrowth of the present analysis allows determining stability of MTT.

The paper is arranged as follows: In the next section, we shall introduce the Einstein equations in $n$-dimensions, and discuss the form of energy-momentum tensor useful for a combined study of dust, perfect fluids, as well as fluids with bulk and shear viscosity. Section \ref{sec3} discusses the various pressureless homogeneous or inhomogeneous 
dust models. We provide evidence for the usefulness of the MTT formulation in identifying the singularity and horizon formation phenomena by scrutinizing a large class of models. Our study 
shows that generically, cosmic censorship holds true. Section \ref{sec4} extends the study to perfect fluids that satisfy certain equations of state. Section \ref{sec5} discusses the gravitational collapse of fluids admitting bulk and shear viscous effects. We show concretely how the
time development of singularity and MTT formation is affected by the viscous parameters. Section \ref{sec6} ends with a discussion on the results.    \\
 
 %%%%%%%%%%%%%%%%%%%%%%%%%%%%%
 %%%%%%%%%%%%%%%%%%%%%%%%%%%

\section{FIELD EQUATIONS FOR SPHERICAL COLLAPSE AND MTT}\label{sec2}
The metric for $n$-dimensional spherical symmetric spacetime with spherical foliation is given by
(we shall use the gravitational units $8\pi G=1$, $c=1$ units in this paper\footnote{The dimensions of the 
$n$-dimensional Newton's constant $G$ is $M^{-1}L^{(n-3)}$. Thus, in these units, $M$ has dimensions of ${L}^{(n-3)}$.})
\begin{equation}\label{ndmetric}
ds^2=-e^{2\alpha(r,t)}dt^2+e^{2\beta(r,t)}dr^2+R^2(r,t)\,d\Omega^2_{\, n-2}\,\,,
\end{equation}
where, the spherical cross section of dimension $(n-2)$ is given by the following form:
\begin{equation}\label{spheremetric}
d\Omega^2_{n-2}= \sum_{i=1}^{n-2} \left[ \prod_{j}^{i-1} \sin^{2} \,\theta^{\,j} \right](d\theta^{\,i})^{\,2}.
\end{equation}
Here, $\alpha(r,t)$, $\beta(r,t)$, and $R(r,t)$ are the three unknown functions to be determined. 
The variable $t$ denotes the time coordinate, $r$ denotes the radial coordinate, and the 
function $R(r,t)$ is radius of the spherical configuration of matter undergoing 
gravitational collapse. For the time being, the matter content of the collapsing cloud is kept unspecified, but as we proceed, we shall involve more intricate terms in the energy-momentum tensor and study 
their effect on gravitational dynamics.
The $(n-2)$ spacelike angular coordinates $\theta^{\,i}$ specify the sphere in Eq. \eqref{spheremetric}. 
We shall consider a general form of the energy-momentum tensor:
\begin{equation}\label{tmunu}
T_{\mu\nu}=(p_{t}+\rho)\,u_{\mu}\,u_{\nu}+p_{t}\,g_{\mu\nu}+(p_{r}-p_{t})\,X_{\mu}\,X_{\nu}
-2\,\eta\,\sigma_{\mu\nu}-\zeta\,\theta\,h_{\mu\nu}\,\,,
\end{equation}
 where $\eta$ and $\zeta$ are the coefficients of shear and bulk viscosity. 
 The quantities $\sigma_{\mu\nu}$, $\theta$, and $h_{\mu\nu}$ are the usual shear tensor, expansion 
 scalar and projection tensor associated with the vector field $u^{\mu}$. The matter variables $\rho$ is the energy density, whereas $p_{t}$ and $p_{r}$ are the tangential and radial components of pressure, respectively. 
 The vector field $X^{\mu}$ is a unit 
radial vector ($X^{\mu}X_{\mu}=1$), whereas $u^{\mu}$ is an unit 
timelike vector field ($u^{\mu}u_{\mu}=-1$) orthogonal to spacelike sections of the metric.
These two vector fields are given by
\begin{eqnarray}\label{ndvectors}
u^{\mu}=e^{-\alpha}(\partial/\partial t)^{\mu},\hspace{0.3cm}
X^{\mu}=e^{-\beta}(\partial/\partial r)^{\mu},
\end{eqnarray}
and naturally, the induced metric on the spacelike sections orthogonal to the $u^{\mu}$ is given by
$h_{\mu\nu}=(g_{\mu\nu}+u_{\mu}\,u_{\nu})$, such that
\begin{eqnarray}\label{exp_shear}
h_{\mu\nu}&=&g_{\mu\nu}+u_{\mu}\,u_{\nu}=e^{2\beta(r,t)}dr^2 + R(r,t)^{2}\,\, d \Omega_{(n-2)}
\end{eqnarray}
The expressions for the quantities in Eq. \eqref{tmunu}, for the metric, Eq. \eqref{ndmetric} are given by
\begin{eqnarray}\label{exp_shear}
\theta&=&\nabla_{\mu}u^{\mu}=e^{-\alpha}\,\left[\dot{\beta}+(n-2)\dot{R}/R\right]\,\\
\sigma^{\mu\nu}&=&\frac{1}{2}(h^{\mu\gamma}\,\nabla_{\gamma}u^{\nu}+h^{\nu\gamma}\,\nabla_{\gamma}u^{\mu})-\frac{1}{(n-1)}\,\theta\,h^{\mu\nu}\label{exp_shear_1}\\ \,
\sigma^{1}{}_{1}&=&\frac{(n-2)}{(n-1)}\,e^{-\alpha}\,\left(\dot{\beta}-\dot{R}/R\right)\,\, ,\\
 {\sigma^{i}}_{j}&=&-\frac{e^{-\alpha}\,}{(n-1)}\,\left(\dot{\beta}-\dot{R}/R\right)\,\delta^{i}{}_{j}\label{exp_shear_2}\, .
\end{eqnarray}
The value of square of shear tensor  $\bar{\sigma}^{2}={\sigma}_{\mu\nu}{\sigma}^{\mu\nu}$ is obtained from Eqs. \eqref{exp_shear_1}-\eqref{exp_shear_2}:
\begin{equation}
\bar{\sigma}^{2}=\frac{(n-2)}{(n-1)}\,e^{-2\alpha}(\dot{\beta}-\dot{R}/R)^{2}.
\end{equation}
We shall use a redefined value $\sigma^{2}= [(n-1)/(n-2)]\bar{\sigma}^{2}$, which is useful to extract the dimensional dependence hidden in various equations and has a value given by
\begin{equation}
\sigma^{2}=\,e^{-2\alpha}(\dot{\beta}-\dot{R}/R)^{2}.
\end{equation}
The components
of the energy-momentum tensor ($T_{\mu\nu}$) in Eq. \eqref{tmunu}
are given by
\begin{eqnarray}
T^{\,0}{}_{0}=-\rho\,,\hspace{0.5cm}
T^{\,1}{}_{1}=p_{r}-\frac{2(n-2)}{(n-1)}\eta\sigma-\zeta\theta\,,\hspace{0.5cm}  
T^{\, i}{}_{j}=\left[p_{t}+\frac{2}{(n-1)}\eta\sigma-\zeta\theta\right]\delta^{i}{}_{j}.
\end{eqnarray}
Given the metric and the energy-momentum tensor, the Einstein tensors are easily obtained and are derived in the Appendix (\ref{sec_71}). The Einstein equations are given by
\begin{eqnarray}\label{grav_coll_eqn}
 \rho(r,t)&=&\frac{(n-2)}{2R^{n-2}R^{\,\prime}}\,F^{\prime}\,(r,t),\\%\hspace{2cm}
 p_{r}(r,t)&=&\bar{\eta}\,\sigma+\zeta\theta-\frac{(n-2)\dot{F}(r,t)}{2R^{n-2}\dot{R}} \,\,,\label{grav_coll_eqn1}\\
 \alpha(r,t)^{\prime}&=& -\frac{(n-2)R^{\, \prime}}{R}\frac{\left(p_{r}-p_{t}-2\eta\sigma \right)}{\left[\rho+p_{r}-\bar{\eta}\,\sigma-\zeta\theta\, \right]}-\frac{\left[\,p_{r}-\bar{\eta}\,\sigma-\zeta\theta \, \right]^{\,\prime}}{\left[\,\rho+p_{r}-\bar{\eta}\,\sigma-\zeta\theta \,\right]}  ,\\
 2{\dot{R}}^{\,\prime}(r,t)&=&R^{\,\prime}\,\frac{\dot{G}}{G}+\dot{R}\,\frac{H^{\,\prime}}{H}\,\,,\\
F(r,t)&=&R^{(n-3)}\left(1-G+H \right)\label{mass_eqn_gen}\, ,
\end{eqnarray}
where the two functions $H(r,t)$, $G(r,t)$, and $\bar{\eta}$ are used for simplification, defined through
$H(r,t)=e^{-2\alpha(r,t)}\,{\dot{R}}^{2}\,$, $\,G(r,t)=e^{-2\beta(r,t)}\,R^{\,\prime\,2}$, and the quantity
$\bar{\eta}=2\,\eta\,[(n-2)/(n-1)]$.

The first two equations above, Eqs. \eqref{grav_coll_eqn} and \eqref{grav_coll_eqn1} are 
the $G_{00}$ and the $G_{11}$ equations, respectively, the third is 
the Bianchi identity, [see Eq. \eqref{alphaprime_eqn_app} in Appendix 1], 
the fourth equation is the $G_{01}$ component, while the fifth equation is 
the equation for mass function [see Eq.\eqref{massfunction_app} in 
the Appendix for a derivation], and defines
the amount of mass contained inside a sphere of radius $R(r,t)$ on a given spacelike slice.
This set of five equations is fundamental in the development of a consistent formalism
of gravitational collapse of a fluid configuration \cite{Goswami:2006ph}. 
Let us count the number of independent field variables.
First, the number of unknowns  are three metric variables $\alpha(r,t)$, $\beta(r,t)$
and $R(r,t)$, the matter variables $\rho(r,t), p_{t}(r,t),\,p_{r}(r,t)$,  
the viscous parameters $\eta$ and $\zeta$, and the mass function $F(r,t)$.
For a choice of $\eta, \zeta$, two free functions remain to be specified on the initial Cauchy slice.
Second, there are usually two ways to reduce the interdependence of 
the variables, (i) to specify the equations of state, relating
the tangential and radial pressures to the matter density, and (ii) subject to 
the energy conditions, specify the initial density and velocity
distribution of the matter configuration on the initial time slice. 
The standard requirement is that initial data must however be regular and smooth, and that the metric functions be at least $C^{2}$. 
Consider, for example, the right side of Eq. \eqref{grav_coll_eqn}. 
The quantity diverges for $R(r,t)=0$, as well as for  $R^{\prime}=0$: 
$R(r,t)=0$ is called a shell focusing 
singularity, whereas the other is called a 
shell-crossing singularity. Shell focusing leads to a 
genuine physical curvature singularity of the spacetime, but shell-crossing 
is not a genuine singularity \cite{Yodzis:1973gha}.
For regularity, we demand that on the initial time
slice $t=t_{i}$, $R(r,t_{i})=r$. On the initial slice, the density function becomes $\rho\sim F^{\prime}/r^{n-2}$.
This implies that, if the density has to be initially regular at the center of the configuration, $F(r,t)\sim r^{n-1}\, m(r,t)$. Here, $m(r,t)$ is a regular 
function of $r$, including at center of the cloud $r=0$, and must at least be  $C^{1}$.
Given such a configuration, the evolution of the matter configuration is carried out by 
the equations of motion in \eqref{grav_coll_eqn},  and results in a unique spacetime.

The following sections contain the details on the formation of trapped regions, that of marginally trapped surfaces, and their time development. So, we briefly discuss the geometry and dynamics of MTT.
Given a $n$-dimensional space $\mathcal{M}$, with metric $g_{\mu\nu}$ of signature $(-,+, +,\cdots)$.
Let $\Delta$ be a hypersurface in $\mathcal{M}$ generated by a future directed vector
field $t^{\mu}$. Let us consider, initially, a 
cross section $S_{0}$ of $\Delta$ with coordinates $\theta_{i}$; see Eq. \eqref{ndmetric}. 
The tangent vector on this foliation is denoted by $t^{\mu}=\partial_{v}$.
We fix the coordinate on $S_{0}$ to be $v_{0}$. Further, we denote the
spatial cross sections that foliate the horizon by $S_{v}$, which
are essentially surfaces of constant $v$. Naturally, if $Q$ is any point on $\Delta$, it has coordinates 
$(v, \theta_{i})$. Given any $S_{0}$, there are two null vector fields orthogonal to it at 
each of its points. Let us denote 
them by $\ell^{\,\mu}$ and $n^{\mu}$. The vector field $t^{\mu}=\left(\ell^{\mu}-C\, n^{\mu}\right)$ 
is a linear combination of these two null vectors and hence is orthogonal to the $(n-2)$-dimensional spacelike foliations of $\Delta$. Since the null vectors are 
normalized through $\ell\cdot n=-1$, we get that $t^{\mu} t_{\mu}=2\,C$; the signature of $t^{\mu}$ is controlled by the scalar $C$. Further, since $t^{\mu}$ is orthogonal to $S_{0}$ and 
tangential to $\Delta$, it generates a foliation preserving flow. The
vector field drags the $S_{0}$ to form the cylinder $\Delta\equiv S_{0}\times \mathbb{R}$. 
This $\Delta$ is called a MTT if the following conditions hold true on each of its foliations \cite{Ashtekar:2005ez}: 
(i) $\theta_{(\ell)}=0$, ~(ii) $\theta_{(n)}<0$. Note that the MTT has no specific signature; it may represent a dynamically evolving spacelike/ timelike horizon (signature positive or negative) or
an isolated (signature null) horizon. In this sense, the MTT formalism is ideally suited to address 
the various phases of a horizon. Also note that, since $t^{\mu}$ is horizon preserving, on $\Delta$,
we must have $\lie_{t}\, \theta_{(\ell)}=0$. This equation gives us (see Appendix 3 for a proof of this property)
\begin{equation}\label{c_eqn}
C=\frac{\lie_{\ell}\, \theta_{(\ell)}}{\lie_{n}\, \theta_{(\ell)}}= \frac{G_{\mu\nu}\ell^{\mu}\,\ell^{\nu} }{\mathcal{R}/2 -G_{\mu\nu}\ell^{\mu}\,n^{\nu}},
\end{equation}
where $G$ is the Einstein tensor, and $\mathcal{R}$ is the Ricci scalar on the spacelike foliations. 
By determining this value, we shall be able to cross-check the nature of MTT in each case. In 
the appendix, (see Sec. \ref{sec73}), the detail calculation of the scalar $C$ is explained for various
class of matter variables. Naturally,
until matter fall in the black hole, the MTT shall be dynamical and must reduce to being isolated 
when no matter falls in. Incidentally, as said earlier, $C$ also determines the stability of black hole horizons \cite{Andersson:2005gq,Andersson:2007fh}. Many examples have been constructed in
\cite{Booth:2005ng,Chatterjee:2008if,Chatterjee:2014jda,Chatterjee:2015fsa,Bengtsson:2008jr,Bengtsson:2010tj,Bengtsson:2013hla, gutti1}. In the following sections, we shall explicitly construct these horizon MTT and comment on their stability.\\

%%%%%%%%%%%%%%%%%%%%%%%%%%%%%%%%%%%%%%%%%%%%%%%%%%%%%%%%%%%%%%%%%%%%%%%%%%%%%%%%
          
\section{SPHERICAL TRAPPED SURFACES IN PRESSURELESS MODELS}\label{sec3}

The Oppenheimer- Snyder- Dutt (OSD) and the Lemaitre- Tolman- Bondi (LTB) models 
describe pressureless collapse scenario where 
the viscosity coefficients $\eta$ and $\zeta$ vanish, and furthermore, we impose $p_{r} = p_{t} =0$, which implies that radial and tangential pressure are equal and vanishing. This leads to 
the following reduced equations from the set of Eqs. \eqref{grav_coll_eqn}-\eqref{mass_eqn_gen}:
\begin{eqnarray}
F^{\prime}&=&\frac{2\rho R' R^{n-2}}{(n-2)}\,\,,\hspace{1.75cm} \dot{F}=0\,\,,\label{p0eqn1}\\
\alpha'&=&0\,, \hspace{3.2cm} \frac{\dot{R}^{\,\prime}}{R^{\,\prime}}=\dot{\beta}\,.\label{p0eqn2}
\end{eqnarray}
The condition $\alpha^{\prime}=0$ implies that $\alpha=\alpha(t)$, and hence, one may redefine the time
coordinate in the metric, Eq. \eqref{ndmetric}. Secondly,
the mass function $\dot{F}=0$ implies that $F=F(r)$. Since $F(r)$ is amount of mass enclosed inside 
the spherical collapsing matter configuration of radius $R(r,t)$, we conclude that the contained mass is time independent. These two features are generically true for all pressureless collapse models in general relativity, independent of spacetime dimensions.
From the second equation in Eq. \eqref{p0eqn2}, we get that $R^{\,\prime}= e^{\beta(r,t)+h(r)}$.
If we redefine $e^{2h(r)}=1-k(r)$, the metric Eq. \eqref{ndmetric} reduces to the following form:
\begin{equation}\label{internal_metric_pressureless}
ds^2=-dt^2+\frac{R^{\,\prime \,2}}{1-k(r)}\,dr^2+R^2(r,t)\,d\Omega^2_{n-2}\,.
\end{equation}
These reductions also affect the time change of $R(r,t)$ as obtained from the mass function Eq. \eqref{mass_eqn_gen}:
\begin{equation}\label{shelleqn_gen}
\dot{R}^{2}=\frac{F(r)}{R^{n-3}}-k(r)\,.
\end{equation}
The function $k(r)$ determines the initial velocity profile of matter, and is used to specify whether
the system is gravitationally bound. First, note from Eq. \eqref{internal_metric_pressureless} 
that for the initial data surface to be spacelike, $k(r)<1$. Second,
as in four dimensions, the Eq. $\eqref{shelleqn_gen}$ may be viewed as an energy equation. $k(r)=0$ implies that the shells 
have zero initial velocity at infinity, signifying that these shells are initially stationary. This value corresponds to marginally bound collapse models.
$k(r)>0$ denotes that the shells are initially infalling (also said to have negative initial velocity).
We shall restrict ourselves to these two gravitationally bound systems.
$k(r)<0$ holds for unbound models, where the initial shells are assumed to have positive (or outward directed)
velocity. For later convenience, we assume that the function $k(r)$ is quadratic in $r$, such that
 $k(r)=K\,r^{2}$. This is a restriction we shall use in this paper. Then, marginally
bound collapse models is for $K=0$, whereas $K=1$ imply bound model. The general 
solution of \eqref{shelleqn_gen} is complicated and depends on the signature of $k(r)$,
particular solutions for choices of $k(r)$ is
studied below.

\subsection{Homogeneous matter configuration}  
Hereafter, the spacetime dimension $n$ is replaced by $(2N+1)$ to simplify the expressions.
For example, the mass function may be written as $F(r)=m\,r^{n-1}=m\, r^{2N}$, where $m$ is a constant 
and independent of $r$. In the following, we shall deal with the marginally bound 
matter configurations, a discussion of bound model is in the appendix
(see Appendix \ref{bound_coll}).
%
%\subsubsection{Marginally bound collapse}
%

For the marginally bound matter configurations at infinity, the differential equation for 
the shell radius, Eq. \eqref{shelleqn_gen}
may be solved to give
\begin{equation}\label{k0timecurve_eqn}
R=r\left[1-N\,\frac{\sqrt{F(r)}}{r^{N}}\,\,t \right]^{\frac{1}{N}}\, .
\end{equation}
 We have assumed $\dot{R}<0$ in the above solution since $R(r,t)$ should decrease with time
 in a gravitational collapse process. 
 This equation [Eq. \eqref{k0timecurve_eqn}] also gives times curve of the collapsing shell.
 The time required for the shell to reach the spacetime singularity is obtained by using $R=0$ 
 in the above Eq. \eqref{k0timecurve_eqn} and is 
 given by $t_{s}=\left(N\,\sqrt{m} \right)^{-1}$. Since $m$ is constant for the homogeneous model,
 this implies that all
 shells reach singularity at the same time for any given  spacetime dimension,
 although the time to reach singularity is dependent on dimensionality.
For simplicity, we shift time coordinate to have $t_{s}=0$. Then, the equation of collapsing shell becomes
\begin{equation}\label{shell_equation_k0_osd}
R(r,t)=\left[N\sqrt{F(r)}\,\,(-t) \right]^{\frac{1}{N}}\,.  
\end{equation}

Note that the metric Eq. \eqref{internal_metric_pressureless} needs to be matched with an 
external solution (see Appendix \ref{sec75} for details about matching requirements). We shall take 
that solution as 
the static Schwarzschild solution with horizon at $R={(2M)}^{\frac{1}{2\,(N-1)}}$. Let the matching 
be carried out on the timelike hypersurface $r=r_{H}$. This implies that for the matching of 
solutions, $F(r_{H})=2M$. Using Eq. \eqref{shell_equation_k0_osd}, the time $t_{H}$ for the shell to reach 
the Schwarzschild radius is
\begin{equation}\label{horizon_reach_time}
t_{H}=-\frac{1}{N}\,F(r_{H})^{\,\frac{1}{2\,(N-1)}}=-\frac{1}{N}\,\left(2M\right)^{\,\frac{1}{2\,(N-1)}}\,.
\end{equation}
\\
Let us now find the motion of the
marginally trapped spheres (MTS) and hence the marginally trapped tube (MTT).
For the spherically symmetric spacetimes, the spherical MTS is defined as the boundary of 
the trapped region and is identical to the condition 
$g^{ab}\,\nabla_{a}R\,\nabla_{b}R=0$. For the metric  Eq. \eqref{ndmetric}, 
this implies $R_{\scaleto{MTT}{4pt}}=F(r)^{1/2(N-1)}$.  
The  time curve of the MTT is obtained from Eq. \eqref{shell_equation_k0_osd} by using
this expression and gives
\begin{equation}\label{ah_trajectory}
R_{\scaleto{MTT}{4pt}}(r,t)=-N\, t.
 \end{equation}
From Eqs. \eqref{ah_trajectory}, and \eqref{horizon_reach_time}, it follows that as 
$R_{\scaleto{MTT}{4pt}}$ equals the horizon radius, $t$ equals $t_{H}$, which implies that at $t=t_{H}$, 
MTT begins to form.\footnote{This feature is more vivid if instead of the equation \eqref{shell_equation_k0_osd}, we had used the original equation \eqref{k0timecurve_eqn}; this equation for the $R_{_{\scaleto{MTT}{4pt}}}$ would become
\begin{equation}
t=-\frac{1}{N}[F(r_{_{\scaleto{MTT}{4pt}}})]^{-1/2}(R^{N}_{_{\scaleto{MTT}{4pt}}}-r^{N}_{_{\scaleto{MTT}{4pt}}}).
\end{equation}
From this equation, it follows that the MTT starts to form when the shell coordinate attains the horizon radius.} Several results follow from these
equations. First, Eq. \eqref{ah_trajectory} shows that the MTT is 
formed at $R=R_{H}$. Second, the radius of MTT reduces with time
and attains $R=0$ exactly at the time of singularity formation $t=t_{s}$. 
Therefore, in short, MTT begins at  $t_{H}$ with radius $R=2M^{1/2(N-1)}$ and shrinks 
at a constant rate $\dot{R}_{\scaleto{MTT}{4pt}}=-N$ to reach singularity at $t=0$. 
This is the unstable MTT that develops inside the collapsing matter. Outside 
the collapsing region, the marginally trapped tube matches with the $R=2M^{1/2(N-1)}$ null surface. 
The trajectory of the MTT is discussed in Figs. \ref{fig:OSDK0_mass}(b) and \ref{fig:OSDK0_mass}(c) below. 

Now we determine the event horizon (EH). The outgoing null ray is given by
$\left(dr/dt \right)_{\scaleto{null}{4pt}}=1/R^{\,\prime}$. Therefore, the rate of change of radius $R(r,t)$ is given by
\begin{equation}\label{eh_radius_equation}
\frac{dR(r,t)}{dt}=\left[R^{\prime}\left(dr/dt \right)_{\scaleto{null}{4pt}}+\dot{R} \right]=(1+\dot{R}).
\end{equation}
We substitute the value of $\dot{R}$ from Eq. \eqref{shell_equation_k0_osd}, and Eq. \eqref{eh_radius_equation} has the following solution:
\begin{equation}\label{radius_eh}
R(t)=\frac{N}{(N-1)}\,t+C_{1}(-t)^{\frac{1}{N}}\,.
\end{equation}
To determine the constant $C_{1}$, we note that the time taken for shell to 
reach $R=(2M)^{\frac{1}{2(N-1)}}$ is $t_{H}=-(1/N)\,\left(2M\right)^{\,\frac{1}{2\,(N-1)}}\,$.
This gives
\begin{equation}
C_{1}=\frac{N^{2}}{(N-1)}\, (-t_{H})^{(N-1)/N}.
\end{equation}
Under these conditions, the null ray is the last outgoing ray and signifies the EH.
Therefore, the time rate of change of the event horizon is obtained from Eq. \eqref{radius_eh}:
\begin{equation}
\dot{R}_{eh}= \frac{N}{(N-1)}\left[1-\, \left(\frac{t_{H}}{t}\right)^{(N-1)/N}\,\right].
\end{equation}
So, clearly, when $t=t_{H}$, the rate of growth of EH stops. The formation time of the EH may be easily obtained 
from the condition that $R_{EH}=0$ at $t=t_{eh}^{i}$, in Eq. \eqref{radius_eh}.  
In the following examples, we use the models from marginally bound collapse scenario to identify the
MTT, spacetime singularity and the horizon. In each case,
we calculate the values of the quantity $C$ using the expression derived in the Appendix \ref{sec73}.     
%%%%%%%%%%%%%%%%%%%%%%%%%%%%%%%%%%%%%%%%%%
\subsubsection*{Examples:}
\begin{figure}[htb]
\begin{subfigure}{.3\textwidth}
\centering
\includegraphics[width=\linewidth]{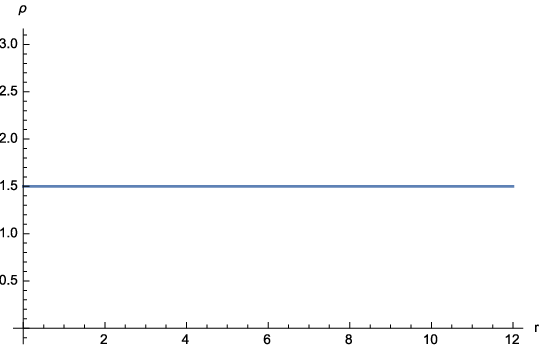}
\caption{}
\end{subfigure}
\begin{subfigure}{.34\textwidth}
\centering
\includegraphics[width=\linewidth]{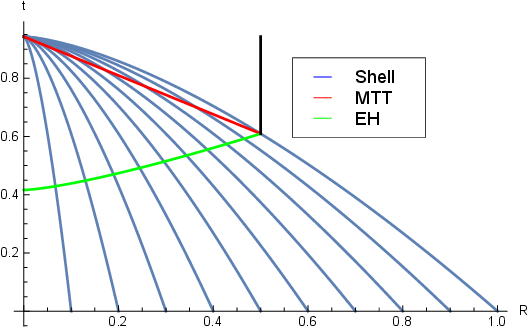}
\caption{}
\end{subfigure}
\begin{subfigure}{.32\textwidth}
\centering
\includegraphics[width=\linewidth]{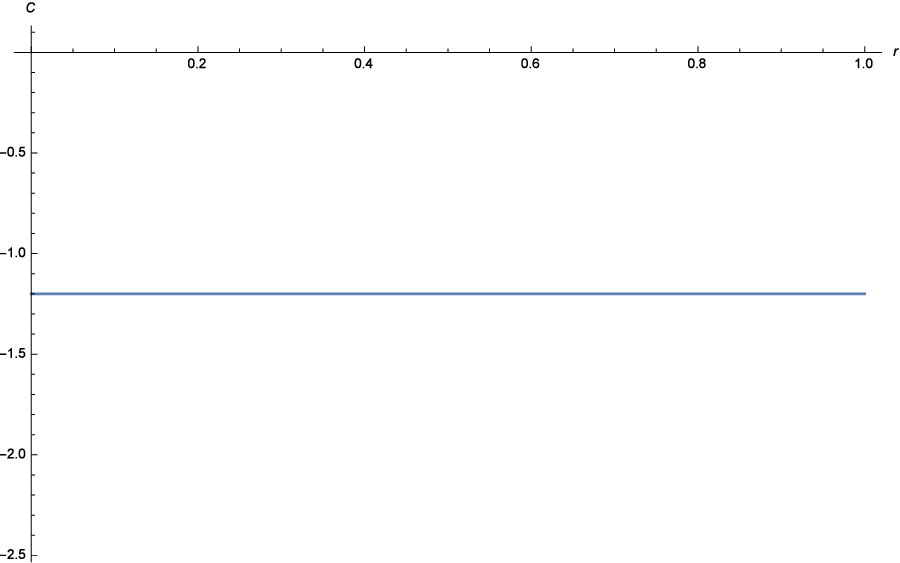}
\caption{}
\end{subfigure}
\caption{Figure (a) gives the plot of density as a function of the shell coordinate $r$, (b) the plot  $R(r,t)$ vs $t$ for the mass profile discussed above, and (c) shows the value of the constant $C$, in four dimensions.
The MTT is timelike since $C$ in the
graph (c) is negative. }
\label{fig:OSDK0_mass}
\end{figure}

We consider the example of a simple mass profile in four dimensions,
$F(r)=mr^{3}$, with $m=(1/2)$, see Fig. \ref{fig:OSDK0_mass}(a).
The motion of the shells is given through the  $t-R$ graph in
Fig. \ref{fig:OSDK0_mass}(b). It is crucial to identify the following features
of collapse process.
First, all shells reach singularity at the same time. 
Secondly, note the teleological behavior of the event horizon:
The last collapsing shell begins at $R(r,t)=1$.
Just as the shell at $R=1$ begins its journey, the event horizon 
anticipates the horizon formation and begins to grow from the center of the cloud.
The shell and the event horizon reach each other exactly at the horizon formation time
or the Schwarzschild radius. 
Third, the MTT also forms at this spacetime point and eventually collapses to singularity. 
Further, the graph of $C$ establishes that $C<0$ throughout,
and hence the MTT is timelike. Here, values of the mass and shell parameters
are chosen to exclude shell-crossing singularities and also to avoid trapped surface
on the initial slice. 

%parameters in the example has been taken in  such a way so as to exclude

For five, and six dimensions, 
the mass function is taken in accordance to $F(r)=mr^{2N}$, with $m=(1/2)$.  
The $t-R(r,t)$ graph is given below in 
Figs. \ref{fig:OSDK0_mass_n5}(a) and \ref{fig:OSDK0_mass_n5}(b). The qualitative features are identical to 
the previous case in four dimensions.
\begin{figure}[htb]
\begin{subfigure}{.35\textwidth}
\centering
\includegraphics[width=\linewidth]{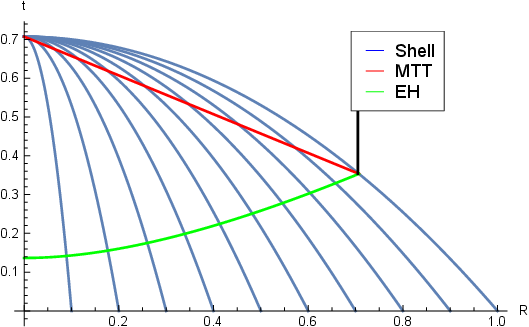}
\caption{}
\end{subfigure}
\begin{subfigure}{.33\textwidth}
\centering
\includegraphics[width=\linewidth]{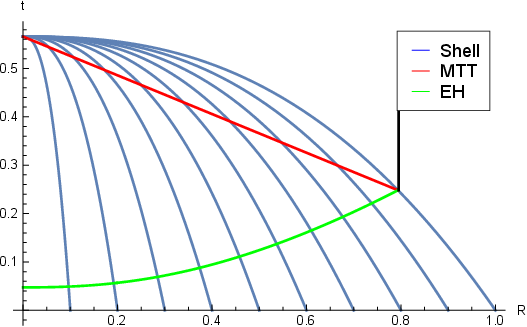}
\caption{}
\end{subfigure}
\begin{subfigure}{.3\textwidth}
\centering
\includegraphics[width=\linewidth]{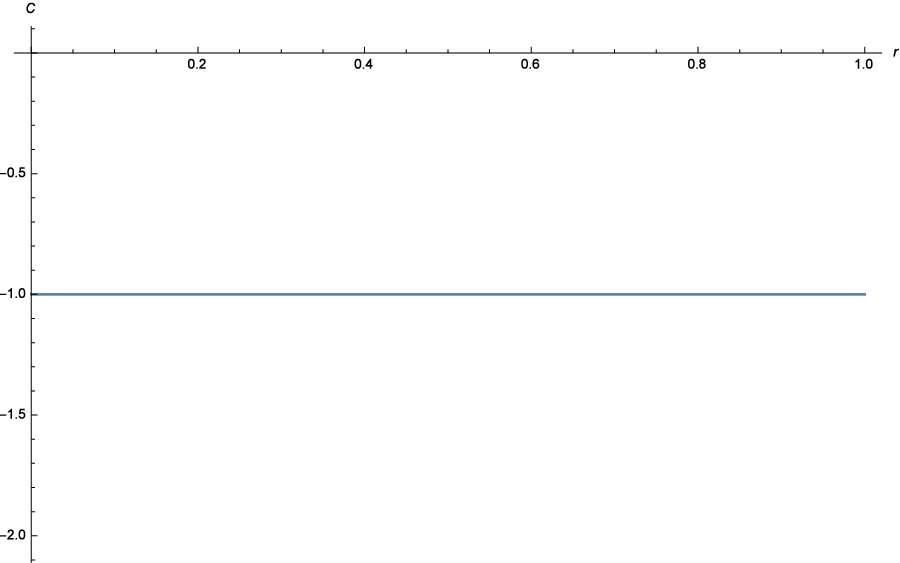}
\caption{}
\end{subfigure}
\caption{Figure (a) gives the plot of $t-R$ for the five dimensions and (b) is for six dimensions. The MTT remains timelike, as seen from the value of $C$ in figure (c). The qualitative feature of the collapse process remains identically characteristic throughout dimensions.}
\label{fig:OSDK0_mass_n5}
\end{figure}
The calculations for bound models of 
gravitational collapse is carried out in the Appendix \eqref{bound_coll}. For completeness, we give the plots 
for bound models Fig. \ref{fig:OSDKg0_mass_n5}. Note that the nature of the graphs are similar to the marginally bound models.
%%%%%%%%%%%%%%%%%
\subsubsection*{Examples for the bound models of collapse:}
These examples concern the bound collapse discussed in the  Appendix \eqref{bound_coll}.
The mass function to be $F(r)=mr^{2N}$, with $m=3$.  
The $t-R(r,t)$ graph is given below in 
Fig. \ref{fig:OSDKg0_mass_n5}(a).  Again, in four dimensions,
all shells reach singularity at $\eta=\pi$; see Eq. \eqref{shell_hb} in the appendix. 
Second, event horizon remains teleological here. Indeed, just as a shell begins to fall at $R(r,t)=2$,
the event horizon begins to grow. As the shell
reaches its Schwarzschild radius at time $\eta=\pi/3$, the event horizon also attains that spacetime point.
The Figures \ref{fig:OSDKg0_mass_n5}(b) and \ref{fig:OSDKg0_mass_n5}(c) are for five and six dimensions. Again, the value of of $C$ is negative, and hence, the MTT is timelike.\\

\begin{figure}[htb]
\begin{subfigure}{.33\textwidth}
\centering
\includegraphics[width=\linewidth]{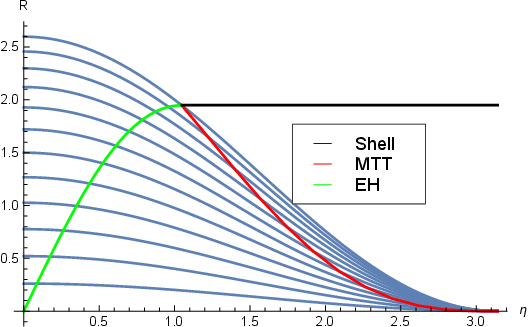}
\caption{}
\end{subfigure}
\begin{subfigure}{.33\textwidth}
\centering
\includegraphics[width=\linewidth]{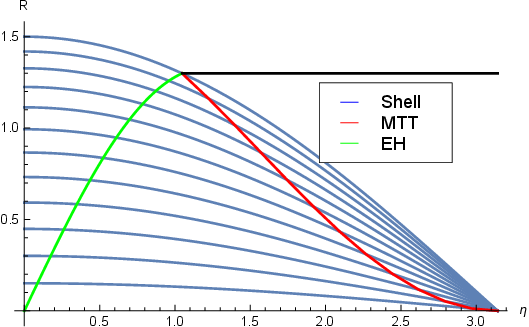}
\caption{}
\end{subfigure}
\begin{subfigure}{.31\textwidth}
\centering
\includegraphics[width=\linewidth]{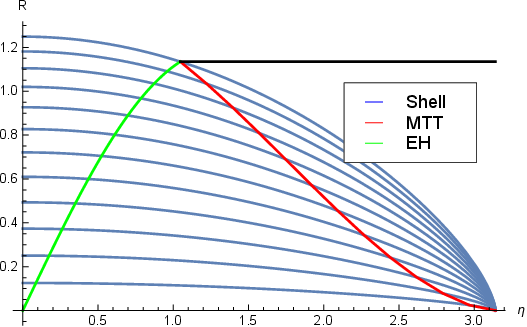}
\caption{}
\end{subfigure}
\caption{Figure (a) gives the plot of $R(r,t)$ vs $t$ for the mass profile discussed above for
(a) four dimensions, (b) five dimensions,(c) six dimensions.}
\label{fig:OSDKg0_mass_n5}
\end{figure}
%
%%%%%%%%%%%%%%%%%%%%%%%%%%%%%%%%%%%%%%%%%%%%%%%%%%%
%%%%%%%%%%%%%%%%%%%%%%%%%%%%%%%%%%%%%%%%%%%%%%%%%%%%%%
%%%%%%%%%%%%%%%%%%%%%%%%%%%%%%%%%%%%%%%%%%%%%%%%%%%%%%
\subsection{Inhomogeneous mass configurations}
For the inhomogeneous collapse models, $F(r,t)=m(r)\,r^{2N}$. Let us first discuss 
the marginally bound models. The solution of the shell equation is same as Eq. \eqref{k0timecurve_eqn}.
The shell begins to collapse at $t_{i}=0$, with $R(t_{i},r)=r$. Time required for the shell to reach
singularity is given by
\begin{equation}
t_{s}(r)=\frac{F(r)^{-1/2}}{N} r^{N} \,.
\end{equation}
Since $F(r)$ is inhomogeneous, shells marked with $r=$constant reach singularity at 
different times. The equation of motion of shells can be rewritten as
\begin{equation}
t=t_{s}(r)-\frac{1}{N}\frac{R^{N}}{\sqrt{F(r)}}\,.
\end{equation}
From this equation, we obtain the time of formation of trapped surface and its radius is obtained using
the condition $R_{\scaleto{MTT}{4pt}}(r,t)=F(r)^{\frac{1}{2(N-1)}}$:
\begin{equation}
t_{\scaleto{MTT}{4pt}}(r)=t_{s}(r)-(1/N)\, {F}^{\frac{1}{2(N-1)}}\,\,;\hspace{1.5cm}
R_{\scaleto{MTT}{4pt}}(r)=N[t_{s}(r)-\,t_{\scaleto{MTT}{4pt}}(r)]\,.  
\end{equation}
\subsubsection*{Example}
In the following, we shall consider the following densities for the spacetimes of dimension $n\equiv (2N+1)$,
\begin{eqnarray}
 \rho_{1}(r) &=&(m/\pi) \left[\frac{2N(2N-1) (2N-2) }{ 2^{2(N-1)} 5^{(2N+1)} }\right]\, (10-r) \Theta (10-r)\nonumber\,,\\
 \rho_{2}(r) &=&(m/\pi)\left[\frac{2N(2N-2)(2N+2) }{2^{(N+5)}\, 5^{N+1}}\right]\left(10-r^2\right) \Theta \left(10-r^2\right),
\end{eqnarray}
The density graphs for $\rho_{1}$ and $\rho_{2}$ are given below in Figs. \ref{fig:LTB_den1_n5}(a) and  \ref{fig:LTB_den2_n6}(a). 
%%%%%%%%%%%%%%%%%%%%%%%%%%%%%%%%%%%%
We shall use the parameter values $m=1$. 
The $t-R(r,t)$ graph for these two density profiles are plotted below for five and six dimensions respectively. Note that in these two cases, the value of $C$ is positive which reflects the fact
that the MTT is spacelike and is evolving with increasing area. These features are clearly demonstrated through the $R-t$ graphs in Figs. \ref{fig:LTB_den1_n5}(b) and  \ref{fig:LTB_den2_n6}(b).

\begin{figure}[htb]
\begin{subfigure}{.33\textwidth}
\centering
\includegraphics[width=\linewidth]{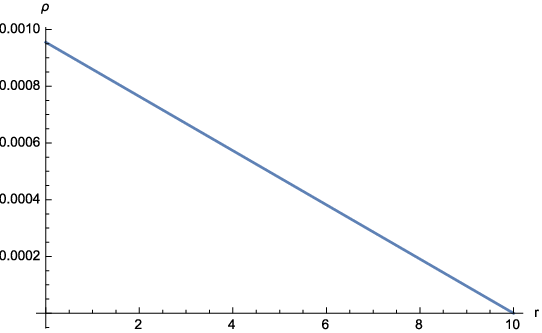}
\caption{}
\end{subfigure}
\begin{subfigure}{.33\textwidth}
\centering
\includegraphics[width=\linewidth]{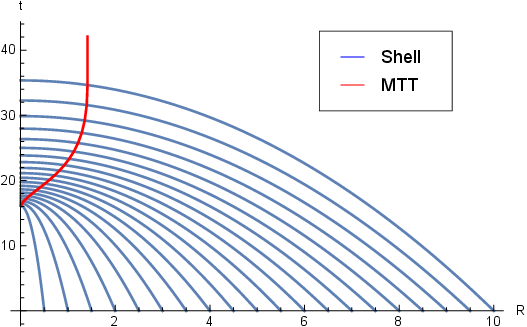}
\caption{}
\end{subfigure}
\begin{subfigure}{.31\textwidth}
\centering
\includegraphics[width=\linewidth]{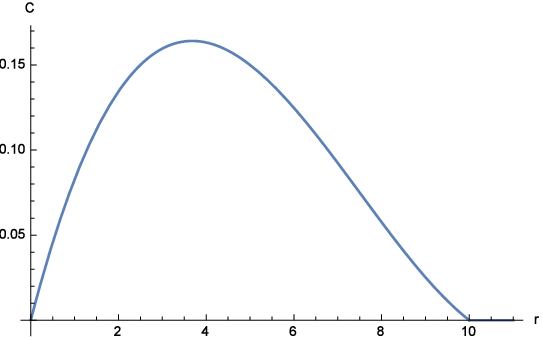}
\caption{}
\end{subfigure}
\caption{Figure (a) gives the density $\rho_{1}$ discussed above, (b) the plot of $R(r,t)$ vs $t$, and (c) the plot of $C$ of MTT in five dimensions. The positive value of $C$ implies that the MTT is spacelike.
When matter stops falling in, the value of $C$ is zero and the MTT attains null signature.}
\label{fig:LTB_den1_n5}
\end{figure}
%

%%%%%%%%%%%%%%%%%%%%
%
\begin{figure}[htb]
\begin{subfigure}{.33\textwidth}
\centering
\includegraphics[width=\linewidth]{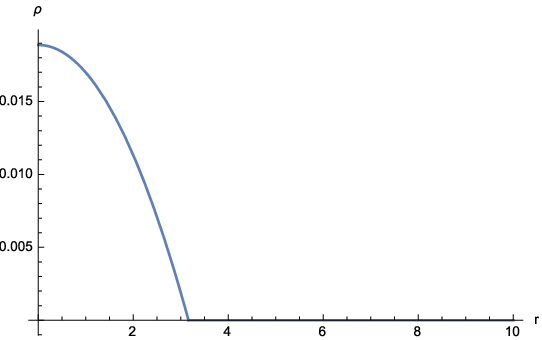}
\caption{}
\end{subfigure}
\begin{subfigure}{.33\textwidth}
\centering
\includegraphics[width=\linewidth]{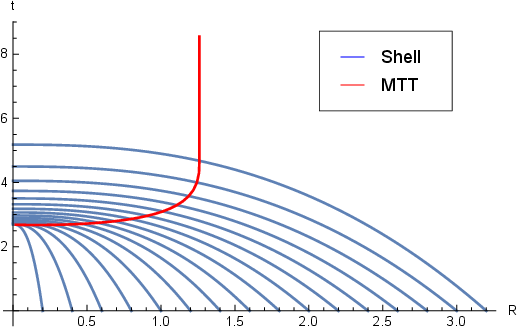}
\caption{}
\end{subfigure}
\begin{subfigure}{.31\textwidth}
\centering
\includegraphics[width=\linewidth]{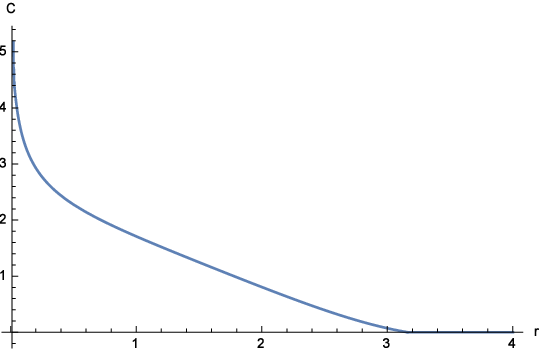}
\caption{}
\end{subfigure}
\caption{Figure (a) gives the density $\rho_{2}$ discussed above, (b) the plot of $R(r,t)$ vs $t$, and (c) the plot of $C$ of MTT in six dimensions show that the MTTs are all spacelike throughout the
collapse process, until mass falls in. The value of $C$ vanishes when the matter stops to fall in. In that situation, the MTT is null and is an isolated horizon. }
\label{fig:LTB_den2_n6}
\end{figure}
\begin{figure}[htb!]
\begin{subfigure}{.33\textwidth}
\centering
\includegraphics[width=\linewidth]{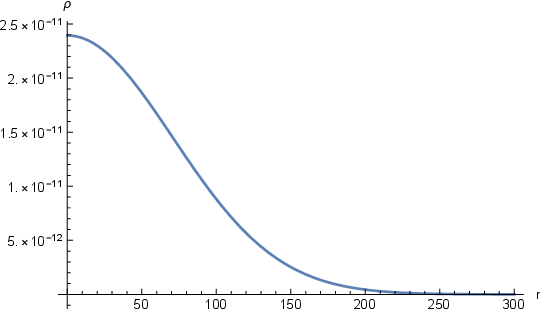}
\caption{}
\end{subfigure}
\begin{subfigure}{.33\textwidth}
\centering
\includegraphics[width=\linewidth]{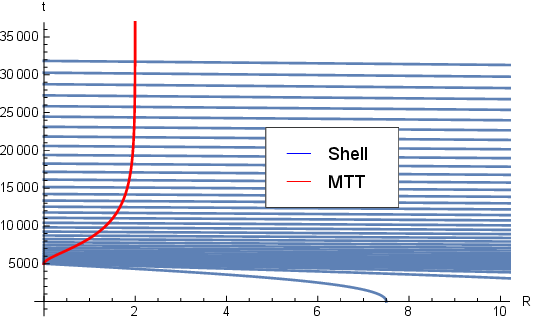}
\caption{}
\end{subfigure}
\begin{subfigure}{.31\textwidth}
\centering
\includegraphics[width=\linewidth]{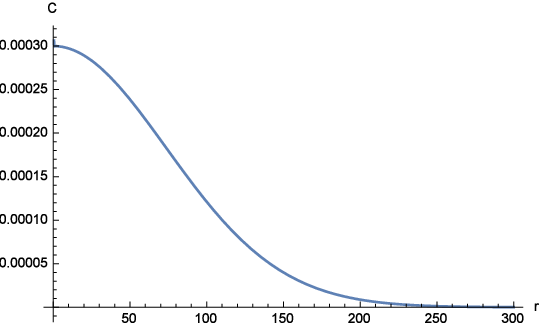}
\caption{}
\end{subfigure}
 \begin{subfigure}{.33\textwidth}
\centering
\includegraphics[width=\linewidth]{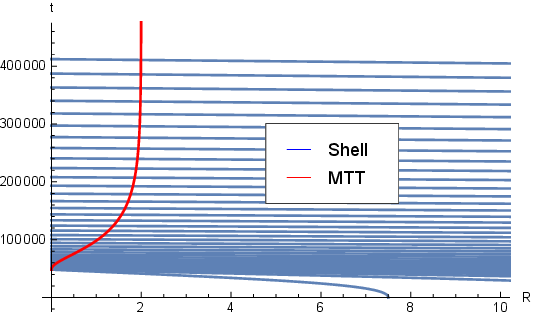}
\caption{}
\end{subfigure}
\begin{subfigure}{.31\textwidth}
\centering
\includegraphics[width=\linewidth]{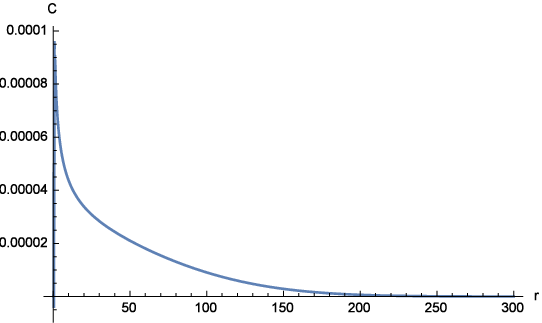}
\caption{}
\end{subfigure}
\caption{Figure (a) gives the Gaussian density discussed above, (b) the plot of $R(r,t)$ vs $t$, and (c) the plot of $C$ of MTT in five dimensions, (d) the  $R(r,t)$ vs $t$ plot, and (e) the plot of $C$ 
for six dimensions. The values of $C$ here shows that the MTT is spacelike (and hence a dynamical horizon) and stable. When no matter falls, that is, when matter density vanishes, as may be seen from graph in (a), the value of $C$ vanishes, and the MTT reaches the isolated horizon phase.}
\label{fig:LTB_gauss_den_n5}
\end{figure}

%
%%%%%%%%%%%%%%%%%%%%%%%%%%%%%%%
%
%\subsubsection*{Bounded models}
For the bound collapse models with $k(r)>0$, the parametric solution of Eq. \eqref{shelleqn_gen} is given by
\begin{eqnarray}\label{inh_bounded_R}
 R(\eta,r)&=&\left[\frac{F(r)}{k(r)}\cos^{2}\left(\eta/2\right) \right]^{\frac{1}{2(N-1)}}	
 \end{eqnarray}
 Using this, one obtains the time curve for shells labelled by a particular value of $r$, and is given by 
 \begin{eqnarray}\label{inh_bounded_t}
 t(\eta,r)&=&-\frac{\left[F(r)/k(r) \right]^{\frac{1}{2(N-1)}}}{N\,\sqrt{k(r)}} \,\left[\cos\left(\eta/2 \right)\right]^{\frac{N}{(N-1)}}\,\,{{}_{2}}\,F_{1}\left[ \frac{1}{2};\,\frac{1}{2}+\frac{1}{2(N-1)};\,\frac{3}{2}+\frac{1}{2(N-1)};\,\cos^{2}\left(\eta/2 \right)\right]\nonumber\\
 &&+\frac{\,\sqrt{\pi}\,\Gamma\left[ (3/2)+\frac{1}{2(N-1)}\right]}{\sqrt{k(r)}\,N\,\,\Gamma\left[1+\frac{1}{2(N-1)} \right]}\,\left[\frac{F(r)}{k(r)} \right]^{\frac{1}{2(N-1)}},
\end{eqnarray}
where ${{}_{2}}\,F_{1}(a;b;c;d)$ is the hypergeometric function. Here, we assume $k(r)=F(r,t)\,r^{-2(N-1)}$ and $F(r,t)=m(r)\,r^{2N}$.  The collapse 
begins at $\eta=0$ and $R(t_{i},r)=r$ and reaches at singularity at $\eta=\pi$ when $R=0$.
Therefore, the time to reach singularity is given by
\begin{equation}
t_{s}(r)=\frac{\,\sqrt{\pi}\,\Gamma\left[ \frac{3}{2}+\frac{1}{2(N-1)}\right]}{N\,\Gamma\left[1+\frac{1}{2(N-1)} \right]}\,\left[\frac{F(r)}{k(r)^{N}} \right]^{\frac{1}{2(N-1)}}
\end{equation}
From the above equation, it is clear that
shells with different initial radius must reach the central singularity at different times. 
From Eq. \eqref{inh_bounded_R}, 
\begin{equation}
\eta=2\cos^{-1}\left[\,\frac{R^{2(N-1)}k(r)}{F(r)}\, \right]^{\frac{1}{2}}.
\end{equation}
Hence, the proper time for the shell to reach the Schwarzschild radius $R=(2M)^{\frac{1}{2(N-1)}}$ is given by
\begin{equation}
\eta_{_{\scaleto{MTT}{4pt}}}=2\cos^{-1}\left[\sqrt{k(r)}\right]\,.
\end{equation}
The equations for the MTT is obtained from 
the condition $R_{\scaleto{MTT}{4pt}}=(2M)^{\frac{1}{2(N-1)}}$. So from Eq. \eqref{inh_bounded_R} and Eq. \eqref{inh_bounded_t}, we have the radius and time curves for the MTT for each fixed value of the shell coordinate:
\begin{eqnarray}
 R_{\scaleto{MTT}{4pt}}(r)&=&r_{_{\scaleto{MTT}{4pt}}}\left[\cos\left(\eta_{_{\scaleto{MTT}{4pt}}}/2 \right) \right]^{\frac{1}{(N-1)}}\,,\\
t_{\scaleto{MTT}{4pt}}(r)&=&-\frac{\left[F(r)/k(r) \right]^{\frac{1}{2(N-1)}}}{N\,\sqrt{k(r)}} \,\left[\cos\left(\eta_{_{\scaleto{MTT}{4pt}}}/2 \right)\right]^{\frac{N}{(N-1)}}\,\,{{}_{2}}\,F_{1}\left[ \frac{1}{2};\,\frac{N}{2(N-1)};\,\frac{3N-2}{2(N-2)};\,\cos^{2}\left(\frac{\eta_{_{\scaleto{MTT}{4pt}}}}{2} \right)\right]\nonumber\\
            &&+\frac{\,\sqrt{\pi}\,\Gamma\left[ \frac{3}{2}+\frac{1}{2(N-1)}\right]}{\sqrt{k(r)}\,N\,\,\Gamma\left[1+\frac{1}{2(N-1)} \right]}\,\left[\frac{F(r)}{k(r)} \right]^{\frac{1}{2(N-1)}}.
        \end{eqnarray}
 % %%%%%%%%%%%%%%%%%%%%%%%%%
 As an example, we consider the mass profile in the form of a Gaussian density distribution:
\begin{eqnarray}\label{gaussian_profile}
 \rho(r) &=&\frac{m_{0}\,(2N-1)}{4\,\pi\, r_{0}^{N}\,\Gamma\left[N \right]}\, e^{-r^{2}/r_{0}^{2}}\,
 \end{eqnarray}
where $r_{0}=100\, m_{0}$, with $m_{0}=1$. The values of these parameters are chosen 
to avoid shell-crossing
singularities and to avoid initial trapped surfaces. 
The $t-R(r,t)$ graph for this density profile is plotted below for $5$, and six dimensions 
respectively, see Figs. \ref{fig:LTB_gauss_den_n5}(b), and  \ref{fig:LTB_gauss_den_n5}(d). Note that in these two cases, the value of $C$ is positive, which reflects the fact
that the MTT is spacelike and evolving with increasing area, see Figs. \ref{fig:LTB_gauss_den_n5}(c), and  \ref{fig:LTB_gauss_den_n5}(e). These features are clearly demonstrated through the $R-t$ graphs in Fig. \ref{fig:LTB_gauss_den_n5}.

%

%%%%%%%%%%%%%%%%%%%%%%%%%%%%%%%%%%%%%%%%%%
%%%%%%%%%%%%%%%%%%%%%%%%%%%%%%%
 
\section{SELF-GRAVITATING PERFECT FLUIDS}\label{sec4}
For the perfect fluids, $p_{r}=p_{t}=p$ and the fluid is viscosity and shear free. Therefore, the Einstein equations in  Eqs. \eqref{grav_coll_eqn}-\eqref{mass_eqn_gen} reduce to\\
\begin{eqnarray}
\rho&=&\frac{(2N-1)F'}{2\,R^{2N-1}\,R'}\,,~~~~p =-\frac{(2N-1)\dot{F}}{2\,R^{2N-1}\,\dot{R}} \,,\\
\alpha'&=& -\frac{p '}{\rho+p } \, , \, ~~~~~~~~\frac{\dot{G}}{G}=2\,\alpha'\,(\,\dot{R}/R'\,)\,,\\
F(r,t)&=&R^{2(N-1)}\left(1-G+H \right) \label{FTD1}\,.
\end{eqnarray}
To solve, we assume an equation of state for the fluid:
$p=k_{p}\,\rho$. The solutions for metric functions are,
\begin{eqnarray}
\exp{(2\alpha)}=\rho^{-2a_{1}}, ~~~~~~~~~\exp{(2\beta)}=\frac{R'^{2}}{1+r^{2}B(r,t)}\,,
\end{eqnarray}
 where $a_{1}=k_{p}/(1+k_{p})$. The metric for this spacetime can be written as
\begin{equation*}
ds^2=-\frac{1}{\rho^{\,2a_{1}}}dt^2+\frac{R'^{2}}{1+r^{2}B(r,t)}\,dr^2+R^2(r,t)\,d\Omega^2_{2N-1}    \,.
\end{equation*}
The equation of motion for the shell is obtained from the $F(r,t)$ equation in Eq. \eqref{FTD1}:
\begin{equation}
\dot{R}=-\rho(r,t)^{\,-a_{1}}\left\{\frac{F(r,t)}{R^{2(N-1)}}+r^{2}B(r,t) \right\}^{\frac{1}{2}}.
\end{equation}
For simplification, let us assume 
the functions are of separable type,
\begin{equation*}
F(r,t)=F_{1}(r)F_{2}(t),\hspace{0.75cm}B(r,t)=B_{1}(r)B_{2}(t), \hspace{0.75cm}\rho(r,t)=\rho_{1}(r)\rho_{2}(t), 
\end{equation*}
In addition to these above choices, we define
\begin{equation}
B_{1}(r)=k(r)/r^{2}, ~~~~~~~~B_{2}(t)=-F_{2}(t)=-\rho_{2}(t)^{2a_{1}},
\end{equation}
along with a parametric form of $R(r,t)$ given by
\begin{equation}
R(\eta,r)=\left[\frac{F_{1}(r)}{k(r)}\,\cos^{2}{\left(\eta/2 \right)} \right]^{\frac{1}{2(N-1)}}\,.
\end{equation}
In terms of the above substitutions and redefinitions, the equation of motion of the collapsing shell relates the variables $t$ and $\eta$ as follows:
\begin{eqnarray}
dt=\frac{\rho_{1}(r)^{\,a_{1}}}{2(N-1)}\left\{\frac{F_{1}(r)}{k(r)^{N}} \right\}^\frac{1}{2(N-1)}\left\{ \cos\left(\eta/2 \right) \right\}^{\frac{1}{(N-1)}}\, d\eta\,.
\end{eqnarray}
The integration of the above equation and subsequent substitution of $\eta$ in terms of $R(r,t)$ gives us
the following relation in terms of the variables $(t,r,R)$: 
\begin{eqnarray*}
 t(\eta,r)&=&({\rho_{1}}^{a_{1}}/N)\,\left(F_{1}/k^N\right)^\frac{1}{2(N-1)}\left[\frac{k\,r^{2(N-1)}}{F_{1}}\right]^{\frac{N}{2(N-1)}}\,{}_{2}F_{1}\left[\frac{1}{2}; \frac{N}{2(N-1)}; \frac{3N-2}{2(N-1)}; \frac{k\,r^{2(N-1)}}{F_{1}} \right]\nonumber\\
 &-&({\rho_{1}}^{a_{1}}/N)\left(F_{1}/k^{N}\right)^\frac{1}{2(N-1)}\left[\frac{k\,R^{2(N-1)}}{F_{1}}\right]^{\frac{N}{2(N-1)}}\,{}_{2}F_{1}\left[\frac{1}{2}; \frac{N}{2(N-1)}; \frac{3N-2}{2(N-1)}; \frac{k\,R^{2(N-1)}}{F_{1}} \right]\,\,.
\end{eqnarray*}
\\
In the following, we take several examples of perfect fluid mass profiles undergoing gravitational collapse.
In all these examples, we choose the mass profile parameters, and the density of state parameter $k_{p}$ carefully so as to avoid the shell-crossing singularities. However, in two of the cases below, [see the examples (b) and (c) below], an initial black hole is present. This is deliberate, and we 
take care not to introduce any other singularities. Also note that, except for the case in example \ref{fig:LTB_large_shell_den_n5}(c), the values of $C$ are all positive as matter falls in the black hole
and the density is not vanishing. This implies that the MTT is spacelike during this time and is called a dynamical horizon. As 
the matter density vanishes, and no matter falls, 
the value of $C$ vanishes, which implies that at that stage the horizon is an isolated horizon.

%%%%%%%%%%%%%%%%%%%%%%%%%%%%%%%%%%%%%%%%%%%%%%%%%%%%%%%%%%
\subsubsection*{Examples:}

\begin{figure}[htb!]
\begin{subfigure}{.42\textwidth}
\centering
\includegraphics[width=\linewidth]{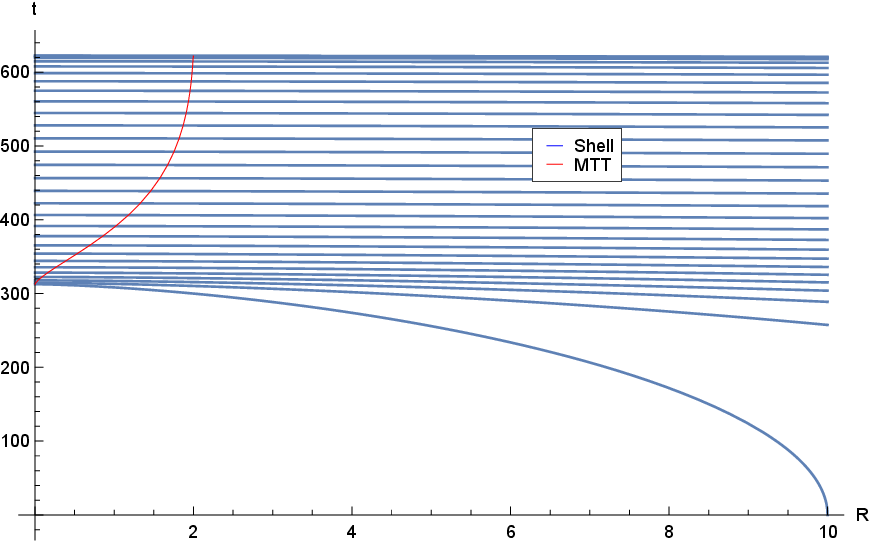}
\caption{}
\end{subfigure}
\begin{subfigure}{.42\textwidth}
\centering
\includegraphics[width=\linewidth]{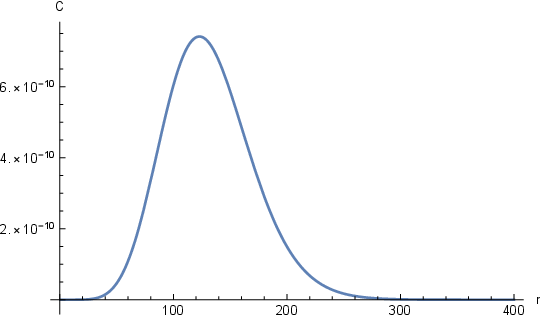}
\caption{}
\end{subfigure}
\begin{subfigure}{.42\textwidth}
\centering
\includegraphics[width=\linewidth]{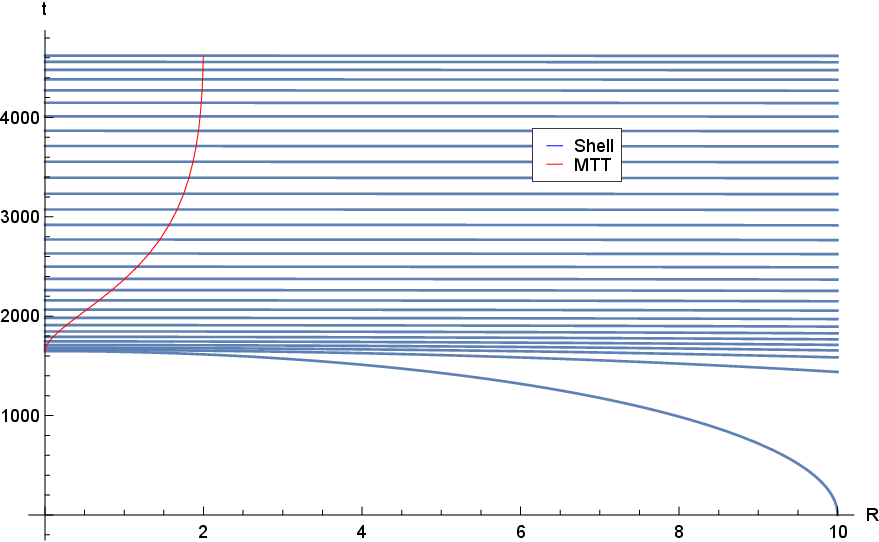}
\caption{}
\end{subfigure}
 \begin{subfigure}{.42\textwidth}
\centering
\includegraphics[width=\linewidth]{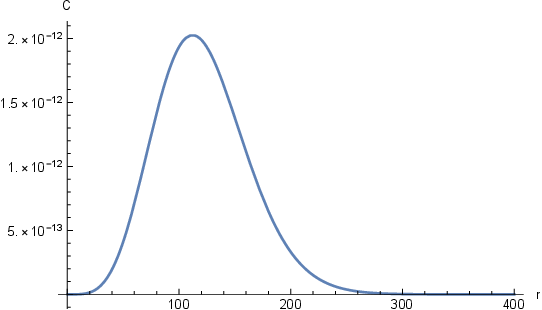}
\caption{}
\end{subfigure}
\caption{Figure (a) gives the plot of shells of radius $R(r,t)$ with time in five dimensions, (b) the plot of $C$ of MTT in five dimensions, (c) the  $R(r,t)$ vs $t$ plot, and (d) the plot of $C$ 
for six dimensions. }
\label{fig:perfect_gauss_den_n5}
\end{figure}

(a) We consider the example of a Gaussian density distribution Eq. \eqref{gaussian_profile} in four and five dimensions undergoing gravitational collapse.
The $t-R(r,t)$ graph for this density profile is plotted below for $k_{p}=0.1$, see Figs. \ref{fig:perfect_gauss_den_n5}(a) and \ref{fig:perfect_gauss_den_n5}(c). 
The value of $C$ is positive implying 
that the MTT is spacelike and evolving with increasing area. These graphs are in 
the Figs. \ref{fig:perfect_gauss_den_n5}(b) and  \ref{fig:perfect_gauss_den_n5}(d).\\
%%%%%%%%%%%

(b) Two shells falling on a black hole of mass: Let us consider the following density profile
\begin{eqnarray}
\rho(r) &=& \frac{m_{0}}{\mathcal{N}r_{0}^{2N}}\exp\left[{\frac{2 \zeta  r}{r_{0}}-\frac{r^2}{{r_{0}}^2}}\right] \left(\frac{r}{r_{0}}-\zeta \right)^2  \,.
 \end{eqnarray}
where $\mathcal{N}$ is a normalization factor, $m_{0}=1/2$, and $r_{0}=10$. The plots for density,
$R(r,t)$ , and $C$ are given below for $k_{p}=0.001$ The parameter choices avoid shell-crossing singularities. The plots of for the density, $R-t$, and $C$ are given in the following Figs. \ref{fig:perfect_maxwell_den_n5}(a), (b), and (c), respectively.\\
\begin{figure}[htb]
\begin{subfigure}{.31\textwidth}
\centering
\includegraphics[width=\linewidth]{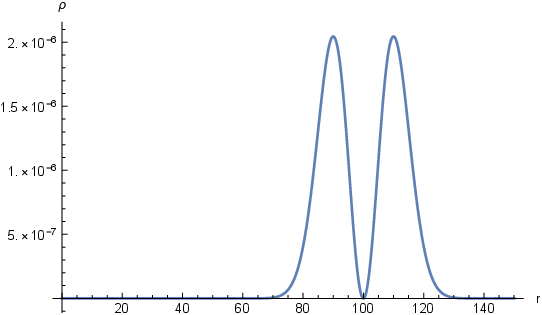}
\caption{}
\end{subfigure}
\begin{subfigure}{.33\textwidth}
\centering
\includegraphics[width=\linewidth]{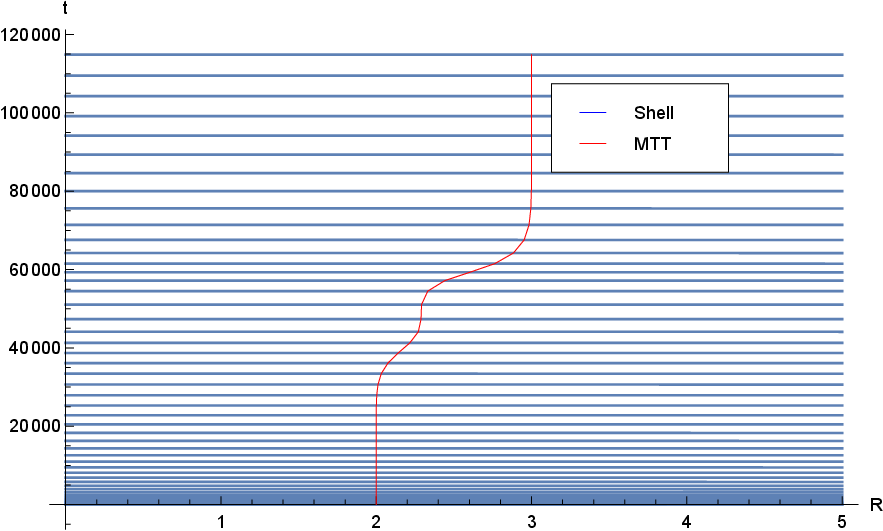}
\caption{}
\end{subfigure}
\begin{subfigure}{.33\textwidth}
\centering
\includegraphics[width=\linewidth]{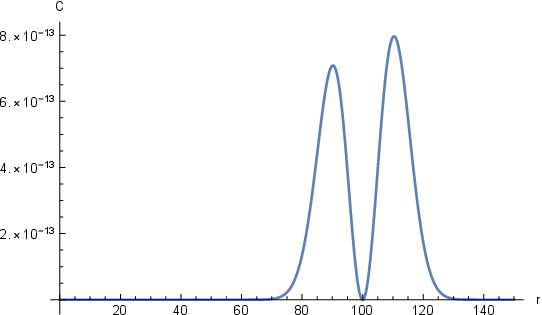}
\caption{}
\end{subfigure}
\caption{Figure (a) shows the density distribution in six dimensions, (b) the plot of $R(r,t)$ vs $t$, and (c) the plot of $C$ in six dimensions.}
\label{fig:perfect_maxwell_den_n5}
\end{figure}
%
%%%%%%%%%%%%%%%%%%%

\begin{figure}[htb]
\begin{subfigure}{.31\textwidth}
\centering
\includegraphics[width=\linewidth]{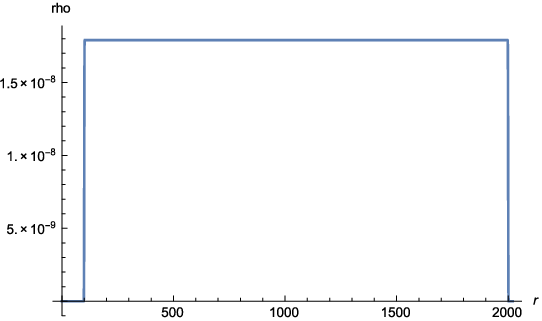}
\caption{}
\end{subfigure}
\begin{subfigure}{.33\textwidth}
\centering
\includegraphics[width=\linewidth]{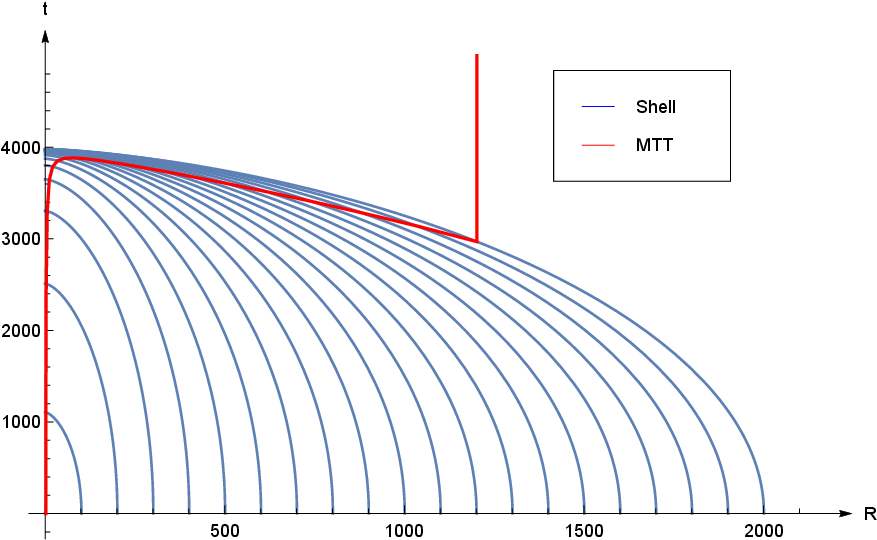}
\caption{}
\end{subfigure}
\begin{subfigure}{.33\textwidth}
\centering
\includegraphics[width=\linewidth]{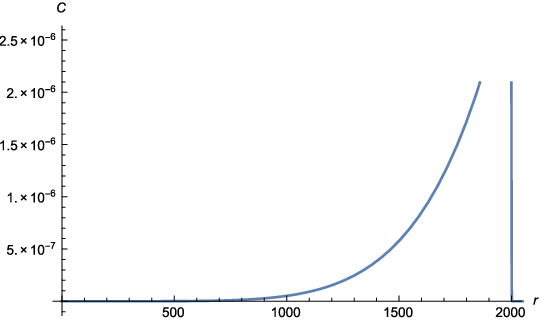}
\caption{}
\end{subfigure}
\caption{Figure (a) shows the density distribution in six dimensions, (b) the plot of $R(r,t)$ vs $t$, and (c) the plot of $C$ of MTT in six dimensions. Note from (b) that the MTT forms at two places: The first one being that of the initial black hole that grows in size, and the second one begins from the last shell (at $r$=2000) and joins with the first MTT. The $C$ however remains spacelike throughout.}
\label{fig:LTB_large_shell_den_n5}
\end{figure}

(c) A large mass falling on a black hole: Let us consider a density
profile of falling on a black hole. 
%The plots for density, $R(r,t)$ , and $C$ are given below for $k_{p}=0.001$ in the Figs. \eqref{fig:LTB_large_shell_den_n5}(a), (b), and (c) respectively. 
%
%
\begin{eqnarray}
\rho(r) &=& \frac{3m_{0}\left(\erf\left[{\frac{r-r_{1}}{M}}\right]-\erf\left[{\frac{r-r_{2}}{M}}\right]\right)}{4\pi(r_{2}-r_{1})(2{r_{1}}^{2}+2r_{1}r_{2}+2{r_{2}}^{2}+3M^{2})}  \,\,.
 \end{eqnarray}
where $M=1$, $m_{0}=600$, $r_{1}=100$ and $r_{2}=2000$. The plots for the density distribution,
collapsing shells $R(r,t)$ vs $t$ , and $C$ are given below for $k_{p}=0.001$ in the Figs. \ref{fig:LTB_large_shell_den_n5}(a), \ref{fig:LTB_large_shell_den_n5}(b), and \ref{fig:LTB_large_shell_den_n5}(c), respectively.
%
%%%%%%%%%%%%%%%%%%%%%%%%%%%%%%%%%%%%%%%%%%

%%%%%%%%%%%%%%%%%%%%%%%%%%%%%%%%%%%%%%%%%%%%%%%%%%%%%%%%%%
%
\section{SPACETIME ADMITTING VISCOUS MATTER FIELDS}\label{sec5}
The mass function has played a fundamental role in our previous discussions, and in higher dimensional
spacetimes with viscous matter, this quantity needs a redefinition.
%Note that in $4$-dimensional spherically symmetric spacetime, mass function is invariantly related to the 
%sectional curvature of the spacelike $2$- dimensional submanifold:
%$F(r,t)=K\, R^{3}$, where $K=R_{\theta\phi\theta\phi}/[R^{4}\sin^{2}\theta]$,
%which is turn, is associated with
To obtain such a quantity, we note that, in four dimensions, one defines the mass function in terms of a NP scalar $\psi_{2}= -(1/2)\,C_{\mu\nu\alpha\beta}\theta^{\mu}\phi^{\nu}\theta^{\alpha}\phi^{\beta}$ \cite{Glass,banerjee}, where $\theta^{\mu}$ and $\phi^{\mu}$ are normalized angular vector fields. The value of $\psi_{2}$ equals $-C_{\theta\phi\theta\phi}/(2\,R^{4}\sin^{2}\theta)$.  In this dimension, one may rewrite this equation as $\psi_{2}=(1/2)e^{-2(\beta+\alpha)}\,C_{trtr}$. Once written in this form, $\psi_{2}$ is independent of the number of angular dimensions and therefore, may be used for spherical systems in arbitrary dimensions. So, in the following, we shall use $\psi_{2}=(1/2)e^{-2(\beta+\alpha)}\,C_{trtr}$ for redefining the Misner- Sharp mass for viscous matter. For the metric Eq. \eqref{ndmetric}, we get
\begin{eqnarray}\label{equation_si2}
 2N(2N-1)\psi_{2}&=&e^{-2\beta}\left[\alpha''+\alpha'^{2}-\alpha'\beta'+\frac{R'^{2}}{R^{2}}-\frac{R''}{R}+\frac{R'\beta'}{R}-\frac{R'\alpha'}{R}\right]-\frac{1}{R^{2}}\nonumber\\
 &&-e^{-2\alpha}\left[\ddot{\beta}+\dot{\beta}^{2}-\dot{\alpha}\dot{\beta}+\frac{\dot{R}^{2}}{R^{2}}-\frac{\ddot{R}}{R}-\frac{\dot{R}\dot{\beta}}{R}+\frac{\dot{R}\dot{\alpha}}{R}\right].
\end{eqnarray}
Using the Einstein equations, Eq. \eqref{equation_si2} may expressed in terms of $F(r,t)$:
\begin{equation}
F(r,t)=\left(\rho+\bar{p}_{t}-\bar{p}_{r}+2\eta\sigma \right)\frac{R^{2N}}{N(2N-1)}-2\psi_{2}R^{2N}\,,
\end{equation}
where $\bar{p}_{r}=(p_{r}-\zeta\theta)$ and 
$\bar{p}_{t}=(p_{t}-\zeta\theta)$. This quantity $\mathscr{F}(r,t)=-\psi_{2}R^{\,2N}$ shall play a role
similar to that of
the mass function. To see this, we find its time and space derivatives:
 \begin{eqnarray}
\mathscr{F}^{\prime}&=&\frac{-R^{\, 2N}\rho^{\,\prime}}{ 2N(2N-1)}-\frac{1}{ 2N(2N-1)}\left[R^{\,2N}\left(\bar{p}_{t}-\bar{p}_{r}+2\eta\sigma \right) \right]'\,,\label{Fprime}\\
\dot{\mathscr{F}}&=&\frac{-1}{2N(2N-1)}\left[R^{\,2N}\left(\rho+\bar{p}_{t}+\frac{\eta\sigma}{N} \right) \right]_{,t}+\frac{R^{\, 2N}}{2N(2N-1)}\left[\bar{p}_{r}-\bar{\eta}\sigma \right]_{,t}\,\label{Fdot}.
\end{eqnarray}
These two equations may be combined to 
extract the time derivative of density $\dot{\rho}$,
\begin{equation}\label{density_dot_eqn}
\dot{\rho}\,e^{-\alpha}+\left[ \rho+ \bar{p}_{r}-\bar{\eta}\sigma \right]\left(\theta-\sigma \right)=0.
\end{equation}
Using the $t$ and $r$ components of the Bianchi identities (see Appendix \ref{sec_71} ) we get
\begin{eqnarray}\label{vis_rho_eqn}
\dot{\rho}&=&-\dot{\beta}\left[\rho+\bar{p_{r}}-\bar{\eta}\sigma  \right]-\frac{(2N-1)\dot{R}}{R}\left[ \rho+\bar{p_{t}}+\frac{\eta\,\sigma}{N} \right] \,,\label{vis_rho_eqn}\\
 {p_{r}}'&=&\left[\,\bar{\eta}\,\sigma \,\right]'+\frac{(2N-1)R'}{R}\left[\bar{p}_{t}-\bar{p}_{r}+2\eta\sigma \right]-\alpha'\left[ \rho+ \bar{p}_{r}-\bar{\eta}\,\sigma \right]\,.\label{vis_rho_eqn2}
\end{eqnarray}
Using the Eq. \eqref{vis_rho_eqn} in Eq. \eqref{density_dot_eqn}, we get an equation relation the matter
variables with the pressure anisotropy $(\bar{p}_{t}-\bar{p}_{r})$ : 
\begin{equation}\label{rhopluspreqn}
\left(\rho+\bar{p}_{r} \right)=2\bar{\eta}\sigma -\bar{\eta}\theta -\frac{(2N-1)\,\dot{R}}{R\,\sigma}\,\left(\bar{p}_{t}-\bar{p}_{r}\right)\,e^{-\alpha}\,.
\end{equation}
This equation shows that
the pressure anisotropy must lead to the generation of 
shear scalar $\sigma$. Using Eq. \eqref{rhopluspreqn} in Eq. \eqref{vis_rho_eqn}, we get
\begin{equation}
\dot{\rho}=\frac{2N(2N-1)\dot{R}^{2}}{R^{2}}\,\left[2\eta+\frac{\left(\bar{p}_{t}-\bar{p}_{r}\right)}{\sigma}\right]\,e^{-\alpha}\,.
\end{equation}
If the density is uniform, $\rho'=0$, simple integration of Eq. \eqref{Fprime} implies that 
\begin{eqnarray}
[2N(2N-1)]\,\mathscr{F}(r,t)=-R^{2N}\left(\bar{p}_{t}-\bar{p}_{r}+2\eta\sigma \right).
\end{eqnarray}
Using this in Eq. \eqref{Fdot} leads to the time evolution equation for $\rho$:
\begin{equation}
\dot{\rho}=-\frac{2N\dot{R}}{R}\left[\,\rho+\bar{p}_{r}-\bar{\eta}\sigma  \,\right]\,.
\end{equation}
Now rewriting Eq. \eqref{vis_rho_eqn}, using the expression of $\theta$ from Eq. \eqref{exp_shear} we get
\begin{equation}
\left(\rho+\bar{p}_{r} \right)\,\theta=-\dot{\rho}\,e^{-\alpha}+\bar{\eta}\sigma^{2}-\frac{(2N-1)\,\dot{R}}{R}\,\left(\bar{p}_{t}-\bar{p}_{r}\right)\,e^{-\alpha} \,\label{rhoprm}.
\end{equation}
This above equation can also be written further as,
\begin{equation}
\dot{\rho}\,e^{-\alpha}=\left[\bar{\eta}\sigma^{2}-(\rho+\bar{p}_{r})\,\theta \right]\left[1-\frac{2N-1}{2N}\frac{\left(\bar{p}_{t}-\bar{p}_{r}\right)}{\left( \rho+\bar{p}_{r}-\bar{\eta}\sigma  \right)} \right]^{-1}\,\label{dotrho}.
\end{equation}
The radial derivative of the Eq. \eqref{rhoprm} along with Eq. \eqref{dotrho} is given by
\begin{eqnarray}
(\rho+\bar{p}_{r})'\,\theta+(\rho+\bar{p}_{r})\,\theta'&=&[\bar{\eta}\sigma^{2}]^{\prime}+\dot{\rho}\,e^{-\alpha}\,\left[\frac{2N-1}{2N}\,\frac{(\bar{p}_{t}-\bar{p}_{r})}{\left( \rho+\bar{p}_{r}-\bar{\eta}\sigma  \right)} \right]'\nonumber\\
&&-(\dot{\rho}\,e^{-\alpha})'\left[1-\frac{2N-1}{2N}\,\,\frac{(\bar{p}_{t}-\bar{p}_{r})}{\left(\rho+\bar{p}_{r}-\bar{\eta}\sigma \right)} \right]\,.
\end{eqnarray}
Again putting value of $\bar{p}_{r}$ from Eq. \eqref{vis_rho_eqn2} in the above equation, we get the final expression relating the geometric parameters to the mater variables:
\begin{eqnarray}
(\rho+\bar{p}_{r})\,\theta'&=&\left[\bar{\eta}\, \sigma^{2} \right]'-\rho^{\, \prime}\,\theta-\theta\,\left[\bar{\eta}\sigma  \right]'-(2N-1)\,\theta\,\frac{R'}{R}\,\left[\bar{p}_{t}-\bar{p}_{r}+2\,\eta\,\sigma \right]\nonumber\\
 &&+\, \alpha'\,\theta\,\left[\rho+\bar{p}_{r}-\bar{\eta}\sigma  \right]+ \frac{2N-1}{2N}\,\left[\frac{(\bar{p}_{t}-\bar{p}_{r})}{\,\left(\rho+\bar{p}_{r}-\bar{\eta}\sigma \right)} \right]^{\,\prime}\,\dot{\rho}\,e^{-\alpha} \nonumber\\
 &&-\,(\dot{\rho}\,e^{-\alpha})'\left[1-\frac{2N-1}{2N}\,\frac{(\bar{p}_{t}-\bar{p}_{r})}{\left( \rho+\bar{p}_{r}-\bar{\eta}\sigma  \right)} \right] \,.
  %+\frac{(n-2)}{(n-1)}\,\left[\frac{\left(\rho+\bar{p}_{r}-\frac{2(n-2)\eta\sigma}{(n-1)}\right)(\bar{p}_{t}-\bar{p}_{r})'-(\bar{p}_{t}-\bar{p}_{r})\left(\rho+\bar{p}_{r}-\frac{2\,(n-2)\,\eta\,\sigma}{(n-1)}\right)}{e^{\alpha}\,\left(\rho+\bar{p}_{r}-\frac{2\,(n-2)\,\eta\,\sigma}{(n-1)}\right)^{2}} \right]
\end{eqnarray}
Note that if  pressure anisotropy vanishes, we may write $\bar{p}_{r}=\bar{p}_{t} \equiv \bar{p}$. If the viscosity parameters $\eta$ and $\zeta$  also vanish, the above equation reduces to
\begin{equation}
(\rho+\bar{p})\,\theta'=0\,.
\end{equation}
This implies that under vanishing viscosity effects and uniform density of perfect fluids, the expansion scalar  must be spatially uniform. The metric is isotropic and conformally flat. However, if the matter admits viscous properties, and in particular has shear viscosity, the spacetime is generically anisotropic. This is an extension of Raychaudhuri's theorem \cite{Raychaudhuri,misra} for $n$-dimensional spacetimes.  

%%%%%%%%%%%%%%%%%%%%%%%%%%%%%%%%%%%%%%%%%%%%%%%%%%%%%%%%%%%%%%%%%%%%%%%%%%%%%%%%%%%%%%%%%%%%%%%%%%%%%%%%%%%%%%%%%%%%%%%%%%%%%%%%%%%%%
\subsection{Spacetimes with time independent matter configurations}
We now construct spacetimes developed due to collapse of viscous fluids.
We assume that matter shells have fixed radial pressure 
$p_{r}=\bar{\eta}\sigma +\zeta\,\theta$. The set of Einstein equations are
\begin{eqnarray}
(2N-1)\,F'&=&2\,\rho R^{2N-1}R^{\,\prime}\,,\\
(2N-1)\,\dot{F}&=&-2R^{2N-1}\dot{R}\left[p_{r}-\bar{\eta}\sigma-\zeta\,\theta \right]=0\,,\\
\alpha'&=&\frac{(2N-1)R'}{\rho R}\left[p_{t}+\frac{\,\eta\,\sigma}{N}-\zeta\,\theta \right]\, \label{TI_alphaprime},\\
 \frac{\dot{G}}{G}&=&2\,\alpha'\frac{\dot{R}}{R'}\,\label{TI_dotG},\\
F(r,t)&=&R^{2(N-1)}\left[1-G+H \right]\,\label{TI_F}.
\end{eqnarray}
Not surprisingly, these equations contain the number of unknowns exceeding the number of independent Einstein equations. The extra information must be supplied to close the system.
These may be given in the form of equations of state $p_{t}=k_{t}\rho$, $\sigma=k_{\sigma}\rho$ and 
$\theta=k_{\theta}\rho$. The solutions of Eqs. \eqref{TI_alphaprime} and \eqref{TI_dotG} becomes
\begin{eqnarray}
\exp{(2\alpha)}&=&R^{2(2N-1)a_{1}} \label{Pot1}\, ,\\
 \exp{(2\beta)}&=&\frac{R'^{2}}{b(r)R^{2(2N-1)a_{1}}}\label{Pot2}\,,
\end{eqnarray}
 where $a_{1}=k_{t}+\,(\eta\, k_{\sigma})/N-\zeta \,k_{\theta}$. The spacetime metric Eq. \eqref{ndmetric}
 becomes
\begin{equation}
ds^2=-R^{2(2N-1)a_{1}}\,dt^2+\frac{R'^{2}}{b(r)\,R^{2(2N-1)a_{1}}}\,dr^2+R^2(r,t)\,d\Omega^2_{n-2}   \,.
\end{equation}
The equation of motion of matter shells is obtained from Eq. \eqref{TI_F}:
\begin{equation} 
\dot{R}=-R^{(2N-1)a_{1}}\left[\frac{F(r)}{R^{2(N-1)}}-1+b(r)\,R^{2(2N-1)a_{1}} \right]^{1/2}\,.
\end{equation}
To simplify, we choose $a_{1}=-(N-1)/(2N-1)$, and the time curve of the collapsing shell is given by
\begin{equation}
dt=-\frac{R^{2(N-1)}\,dR}{\left[F(r)+b(r)-R(r,t)^{2(N-1)} \right]^{1/2}}\,\label{TI_dt}.
\end{equation}
 To solve the integral, we choose a parametric form to relate function $R(r,t), F(r)$ and $b(r)$. A simple choice is given by
\begin{equation}
R(\eta,r)=\left[\left\{F(r)/b(r) \right\}\cos^{2}\left(\eta/2\right) \right]^{\frac{1}{2(N-1)}}\,.
\end{equation}
Using this form the equation of collapsing shell Eq. \eqref{TI_dt} simplifies,
\begin{equation}
dt=\frac{1}{4(N-1)}\left[F(r)/b(r) \right]^{\frac{(2N-1)}{2(N-1)}}\frac{\sin{\eta}\, \
\cos{\left(\eta/2 \right)}^{\frac{1}{(N-1)}}}{\left[F+b-(F/b)\cos^{2}\left(\eta/2 \right) \right]^{\frac{1}{2}}} d\eta\,.
\end{equation}
The solution of the above equation is given by
\begin{eqnarray}
t(\eta,r)&=&\left[\frac{\left\{F(r)/b(r)\right\}^{\frac{(2N-1)}{2(N-1)}}}{(2N-1)\sqrt{F(r)+b(r)}}\right]\,{{}_{2}}\,F_{1}\left[\frac{1}{2};\, \frac{2N-1}{2(N-1)};\, \frac{4N-3}{2(N-1)};\, \frac{F(r)}{b(r)\,\left(F(r)+b(r)\right)} \right]\nonumber\\
 &-&\left\{\frac{\left\{F(r)/b(r)\,\cos^{2}{\eta/2} \right\}^{\frac{(2N-1)}{2(N-1)}}}{(2N-1)\sqrt{F(r)+b(r)}}\right\} \,{{}_{2}}\,F_{1}\left[\frac{1}{2};\, \frac{2N-1}{2(N-1)};\, \frac{4N-3}{2(N-1)};\, \frac{F(r)\cos^{2}{ \frac{\eta}{2}}}{b(r)\,\left(F(r)+b(r)\right)} \right]\,.\nonumber\\
%  t(\eta,r)&=&-\frac{\sqrt{b(r)}\,\left(\frac{F(r)}{b(r)} \right)^{\frac{(n-2)}{2(N-1)}}\,\sqrt{F(r)+b(r)-\frac{F(r)}{b(r)}}\,\,{{}_{2}}\,F_{1}\left(\frac{1}{2}, \frac{n-2}{2(N-1)};\, \frac{2\,n-5}{2(N-1)};\, \frac{F(r)}{b(r)\,\left(F(r)+b(r)\right)} \right)}{(n-2)\sqrt{F(r)+b(r)}\,\sqrt{{b(r)(F(r)+b(r))-F(r)}}}\nonumber\\
 %           &&-\frac{\sqrt{b+F}\left(\frac{F}{b} \right)^{\frac{n-2}{2(N-1)}}\sqrt{\frac{b R^{2(N-1)}}{F}}^{\frac{2(n-2)}{2(N-1)}}{{}_{2}}\,F_{1}\left(\frac{1}{2}, \frac{n-2}{2(N-1)}; \frac{2n-5}{2(N-1)}; \frac{R^{2(N-1)}}{b+F} \right)}{(n-2)(b+F)}            
 \end{eqnarray}
We use this equation to study the following examples, and in each case, we assume that the function $b(r)=F(r)\, r^{-2(N-1)}$.
%
%%%%%%%%%%%%%%%%%%%%%%%%%%%%%%%%%%%%%%%%%%%%%%%%%%%%
\subsubsection*{Examples}
(a) Let us consider the following density profile
to describe the density distribution of a viscous fluid. This density falls on an already formed black hole of mass $M$.
\begin{eqnarray}
\rho(r) &=& \frac{m_{0}}{\mathcal{N}r_{0}^{2N}}\exp\left[{\frac{2 \zeta  r}{r_{0}}-\frac{r^2}{{r_{0}}^2}}\right] \left(\frac{r}{r_{0}}-\zeta \right)^2  \,,
 \end{eqnarray}
where $\mathcal{N}$ is a normalization factor, $m_{0}=M/2$, and $r_{0}=10$. The plots for density,
$R(r,t)$ , and $C$ are given below in Figs. \ref{fig:TI_maxwell_den_n5}(a), \ref{fig:TI_maxwell_den_n5}(b), and \ref{fig:TI_maxwell_den_n5}(c), respectively.
\begin{figure}[htb]
\begin{subfigure}{.33\textwidth}
\centering
\includegraphics[width=\linewidth]{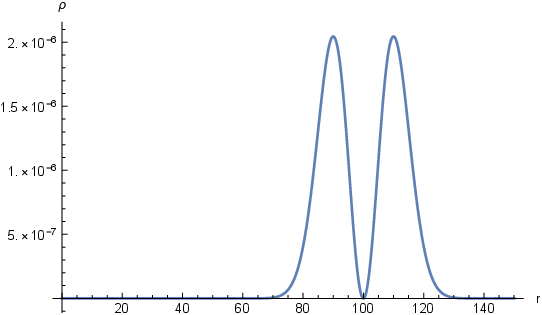}
\caption{}
\end{subfigure}
\begin{subfigure}{.33\textwidth}
\centering
\includegraphics[width=\linewidth]{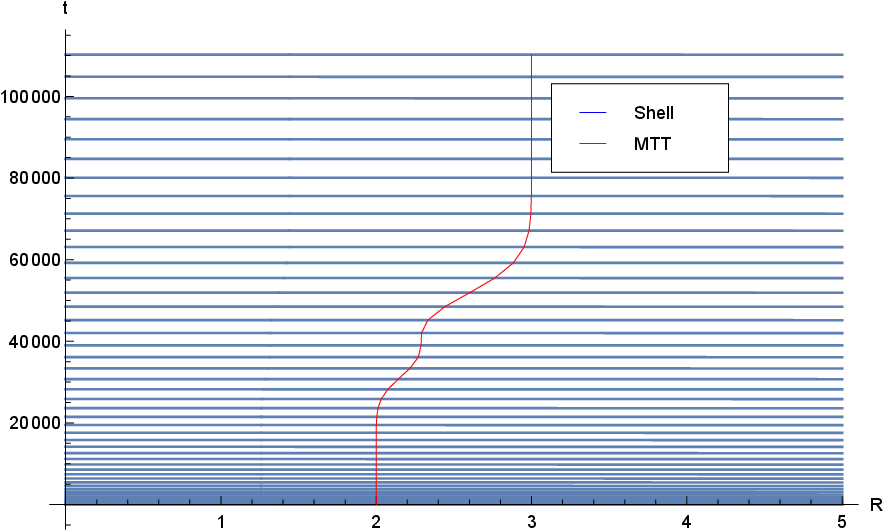}
\caption{}
\end{subfigure}
\begin{subfigure}{.31\textwidth}
\centering
\includegraphics[width=\linewidth]{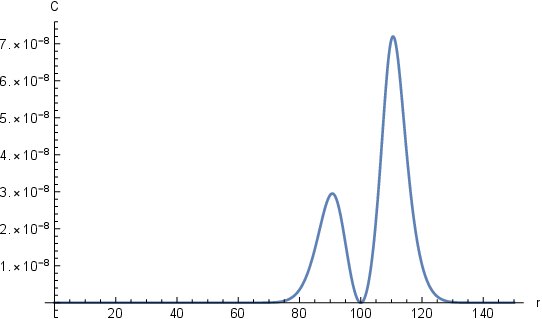}
\caption{}
\end{subfigure}
\caption{Figure (a) shows the density distribution, (b) collapsing shells, and (c) the plot of $C$ in six dimensions.}
\label{fig:TI_maxwell_den_n5}
\end{figure}
%

%%%%%%%%%%%%%%%%%%%%

%

(b) The density distribution is given by the following form:
\begin{equation}
\rho_{2}(r) =\frac{m_{0}(2N-1)}{8\pi\, r_{0}^{2N} \Gamma\left[2N\right]}\,e^{-\frac{r}{r_{0}}},
\end{equation}
where $r_{0}=100\, m_{0}$, with $m_{0}=1$. The plots for density,
collapsing shells, and $C$ are given below in Fig. \ref{fig:TI_Exp_den_n5}.
\begin{figure}[htb]
\begin{subfigure}{.33\textwidth}
\centering
\includegraphics[width=\linewidth]{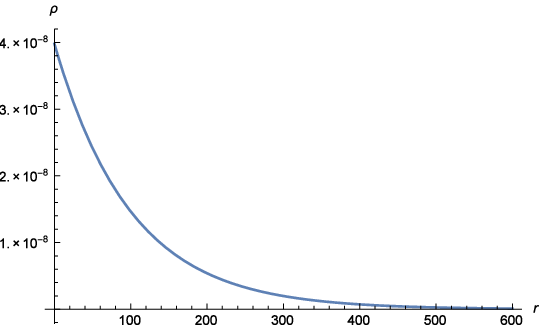}
\caption{}
\end{subfigure}
\begin{subfigure}{.33\textwidth}
\centering
\includegraphics[width=\linewidth]{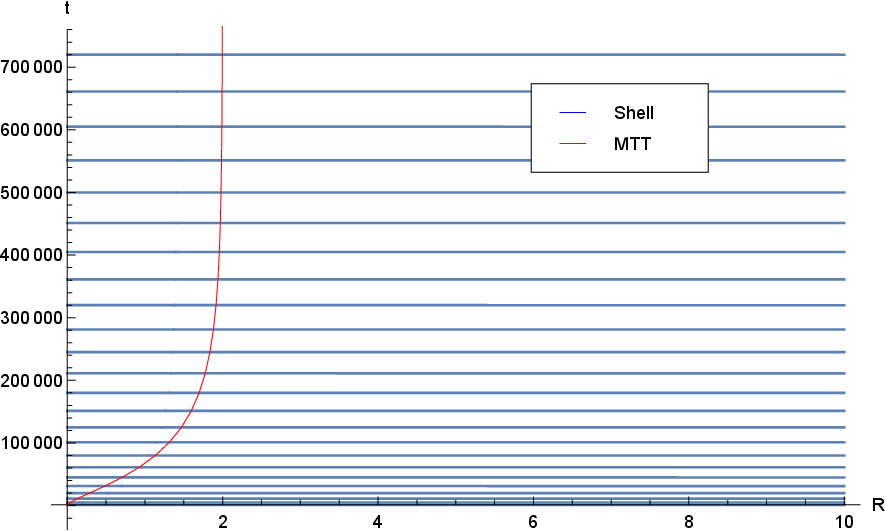}
\caption{}
\end{subfigure}
\begin{subfigure}{.31\textwidth}
\centering
\includegraphics[width=\linewidth]{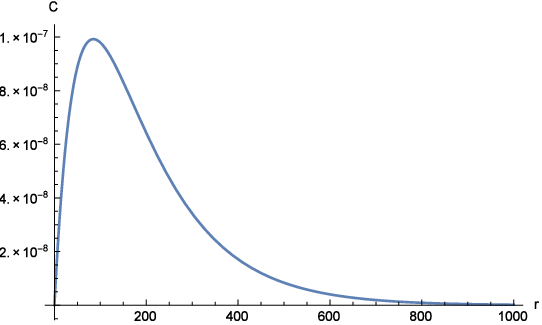}
\caption{}
\end{subfigure}
\caption{Figure (a) shows the density distribution (b) the plot of collapsing shells, and (c) the plot of $C$ of MTT in five dimensions.}
\label{fig:TI_Exp_den_n5}
\end{figure}
%

%%%%%%%%%%%%%%%%%%%%

 %       
\subsection{Spacetimes admitting matter with time dependent mass function}
This is the complete system of matter and gravity in full generality, with the mass function being time dependent as well. The Einstein equations are then
\begin{eqnarray}
\rho&=&\frac{(2N-1)F'}{2\,R^{2N-1}\,R'}\,,\\
p_{r}&=&\bar{\eta}\sigma+\zeta\,\theta-\frac{(2N-1)\dot{F}}{2\,R^{2N-1}\,\dot{R}} \,,\\
\alpha'&=& -\frac{(2N-1)\,R'}{R}\frac{\left[p_{r}-p_{t}-2\,\eta\,\sigma \right]}{\left[\rho+p_{r}-\bar{\eta}\sigma-\zeta\,\theta \right]}-\frac{\left[p_{r}-\bar{\eta}\sigma-\zeta\,\theta \right]'}{\left[\rho+p_{r}-\bar{\eta}\sigma-\zeta\theta \right]}  \,,\\
\frac{\dot{G}}{G}&=&2\,\alpha'\,\frac{\dot{R}}{R'}\,,\\
F(r,t)&=&R^{2(N-1)}\left(1-G+H \right) \label{FTD}\,.
\end{eqnarray}
Note that the equation for radial pressure has a nonvanishing $\dot{F}(r,t)$.
We assume a set of constraints on
the dynamical quantities, $p_{t}=k_{t}\,\rho$, $\sigma=k_{\sigma}\,\rho$, $\theta=k_{\theta}\,\rho$ and $p_{r}=k_{r}\,\rho$. Using these conditions, the solutions of metric functions are
\begin{eqnarray}
\exp{(2\alpha)}&=&\frac{R^{2(2N-1)a_{1}}}{\rho^{2a_{2}}},  ~~\exp{(2\beta)}=\frac{R'^{2}}{1+r^{2}B(r,t)}\,,
\end{eqnarray}
 where the parameters $a_{1}$ \& $a_{2}$ are defined as
 \begin{equation}
 a_{1}=\frac{\left( k_{t}-k_{r}+2\eta k_{\sigma}\right)}{\left(1+k_{r}-\bar{\eta}k_{\sigma}-\zeta k_{\theta} \right)}, ~~~~~
 a_{2}=\frac{\left(k_{r}-\bar{\eta}k_{\sigma}-\zeta k_{\theta} \right)}{\left(1+k_{r}-\bar{\eta}k_{\sigma}-\zeta k_{\theta} \right)}.
 \end{equation}
 The metric for this spacetime, Eq. \eqref{ndmetric} becomes
\begin{equation}
ds^2=-\frac{R^{2(2N-1)a_{1}}}{\rho^{2a_{2}}}dt^2+\frac{R'^{2}}{1+r^{2}B(r,t)}\,dr^2+R^2(r,t)\,d\Omega^2_{n-2}    \,.
\end{equation}
 The equation of motion of the matter shells is obtained from the mass function Eq. \eqref{FTD}
 \begin{equation}
\dot{R}=-R^{(2N-1)a_{1}}\rho(r,t)^{\,-a_{2}}\left\{\frac{F(r,t)}{R^{2(N-1)}}+r^{2}B(r,t) \right\}^{\frac{1}{2}}.
 \end{equation}
For simplification of above equation, let us assume the functions in the above equation are of separable type,
 \begin{equation}
F(r,t)=F_{1}(r)F_{2}(t),\hspace{0.75cm}B(r,t)=B_{1}(r)B_{2}(t), \hspace{0.75cm}\rho(r,t)=\rho_{1}(r)\rho_{2}(t).
\end{equation}
In addition to these above choices, for simplification, we choose
some parameters having form, $B_{1}(r)=k(r)/r^{2}$, $B_{2}(t)=-F_{2}(t)=-\rho_{2}(t)^{2a_{2}}$ and parametric form of $R(r,t)$ is given by,
\begin{equation}
 R(\eta,r)=\left[\left(\frac{F_{1}(r)}{k(r)} \right)\cos^{2}{\left(\frac{\eta}{2} \right)} \right]^{\frac{1}{2(N-1)}}\,.
 \end{equation}
The equation of motion of the collapsing shell is given by
 \begin{eqnarray}
 dt=\frac{\rho_{1}(r)^{\,a_{2}}}{2(N-1)\sqrt{k(r)}}\left\{\frac{F_{1}(r)}{k(r)} \right\}^{\left\{\frac{1-(2N-1)a_{1}}{2(N-1)} \right\}}\left\{ \cos\left(\frac{\eta}{2} \right) \right\}^{\frac{2}{2(N-1)}\left\{1-a_{1}(2N-1) \right\}} d\eta\,.
 \end{eqnarray}
The solution of the above equation is given by
\begin{eqnarray}
 t(\eta,r)&=&\,\frac{2\,\rho_{1}(r)^{a_{2}}\left\{\cos{\left(\eta/2\right)} \right\}^{\frac{(2N)-2\,a_{1}(2N-1)}{2(N-1)}} {{}_{2}}\,F_{1}\left[\frac{1}{2}; \frac{(2N)-2a_{1}(2N-1)}{4(N-1)}; \frac{2a_{1}(2N-1)-6N+4}{4-4N}; \cos^{2}{\left(\eta/2\right)} \right]}{\{2a_{1}(2N-1)-(2N)\}\,k(r)^{\frac{(2N)-2a_{1}(2N-1)}{4(N-1)}}\,{F_{1}(r)}^{\frac{a_{1}(2N-1)-1}{2(N-1)}}}\nonumber\\
       &-&\frac{2\,\rho_{1}(r)^{a_{2}}\, {{}_{2}}\,F_{1}\left[\frac{1}{2}; \frac{(2N)-2a_{1}(2N-1)}{4(N-1)}; \frac{2a_{1}(2N-1)-6N+4}{4-4n}; 1 \right]}{\{2a_{1}(2N-1)-(2N)\}\,k(r)^{\frac{(2N)-2a_{1}(2N-1)}{4(N-1)}}\,{F_{1}(r)}^{\frac{a_{1}(2N-1)-1}{2(N-1)}}}\,\,.
%t(\eta,r)&=&\frac{2\rho_{1}^{a_{2}}\left(\frac{F_{1}(r)}{k(r)} \right)^{\frac{1-a_{1}(n-2)}{2(N-1)}}\left(\frac{k(r)R^{2(N-1)}}{F_{1}(r)} \right)^{\frac{-2a_{1}(n-2)+n-1}{22(N-1)}} {{}_{2}}\,F_{1}\left(\frac{1}{2}, \frac{-2a_{1}(n-2)+n-1}{22(N-1)}; \frac{2a_{1}(n-2)-3n+7}{6-2n}; \frac{k(r)R^{2(N-1)}}{F_{1}} \right)}{\sqrt{k(r)}(2a_{1}(n-2)-n+1)}\nonumber\\
 %&&-\frac{2\rho_{1}^{a_{2}}{{}_{2}}\,F_{1}\left(\frac{1}{2}, \frac{-2a_{1}(n-2)+n-1}{22(N-1)}; \frac{2a_{1}(n-2)-3n+7}{6-2n};1 \right)\left(r^{2(N-1)} \right)^{\frac{1-a_{1}(n-2)}{2(N-1)}}}{(2a_{1}(n-2)-n+1)\sqrt{F_{1}r^{3-n}}} 
\end{eqnarray}
\subsection*{Example}
\begin{figure}[htb!]
 \begin{subfigure}{.42\textwidth}
\centering
\includegraphics[width=\linewidth]{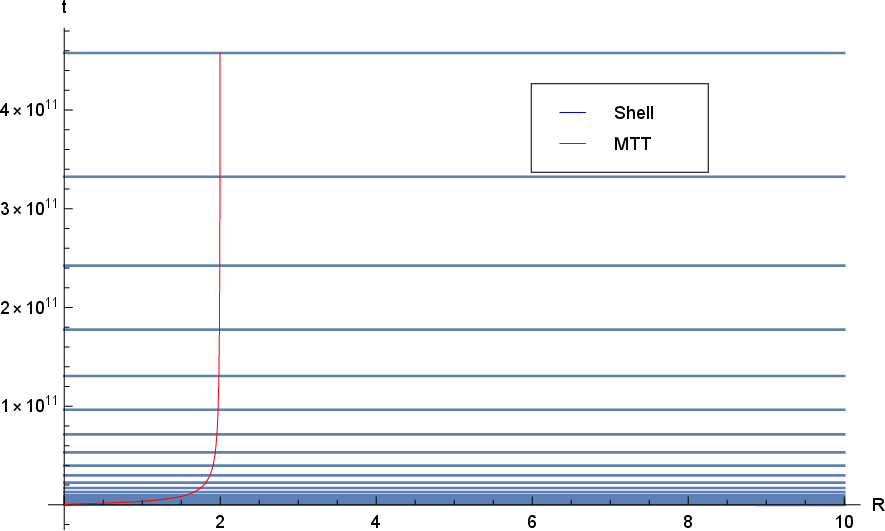}
\caption{}
\end{subfigure}
\begin{subfigure}{.42\textwidth}
\centering
\includegraphics[width=\linewidth]{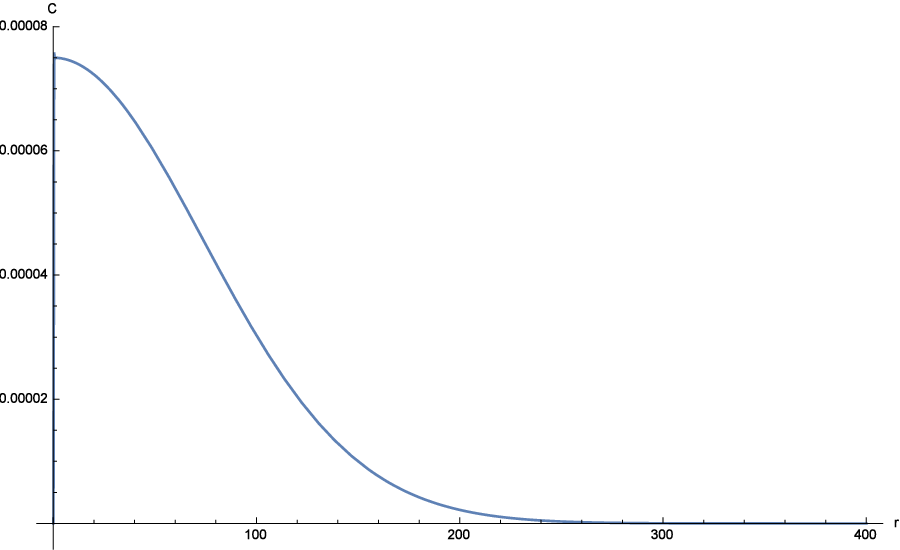}
\caption{}
\end{subfigure}
\begin{subfigure}{.42\textwidth}
\centering
\includegraphics[width=\linewidth]{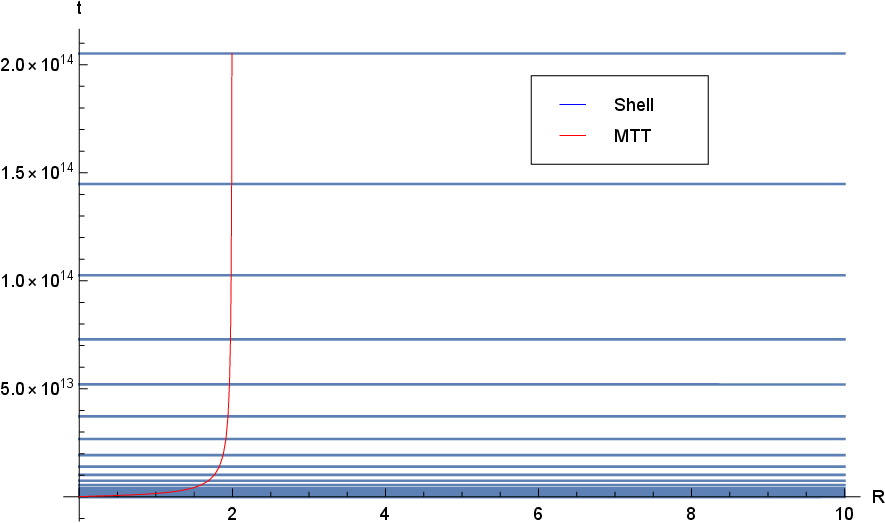}
\caption{}
\end{subfigure}
\begin{subfigure}{.42\textwidth}
\centering
\includegraphics[width=\linewidth]{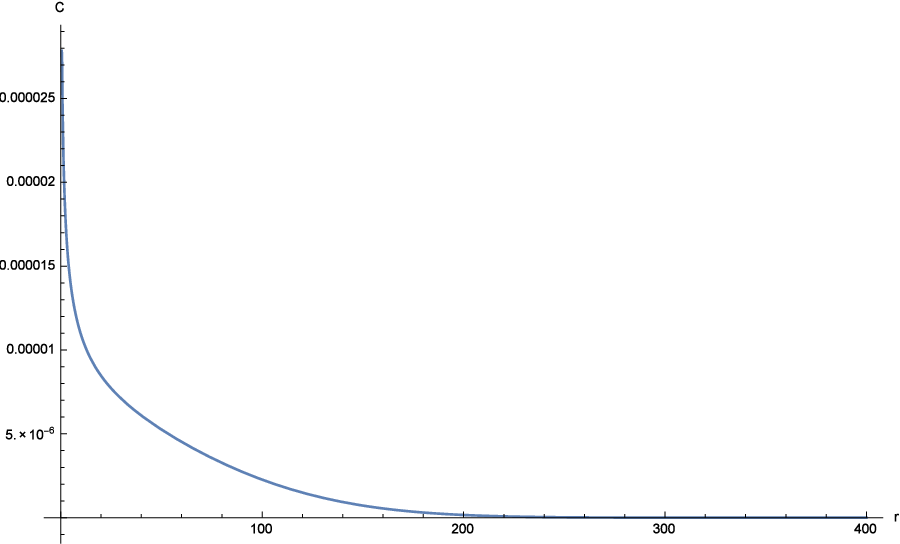}
\caption{}
\end{subfigure}
\caption{Figure (a) shows the collapsing shells, and (b) the plot of $C$ in five dimensions, (c) collapsing shells, and (e) the plot of $C$ 
for six dimensions. The choices of parameters
are same as in the earlier cases and do not allow shell-crossing singularities or trapped surfaces on the initial data surface. }
\label{fig:Gen_gauss_den_n5}
\end{figure}

In the following we shall consider the Gaussian density profile considered earlier in Eq. \eqref{gaussian_profile}. The plots for shells, evolution of MTT, and its signature for 
five and six dimensions are given in Fig. \eqref{fig:Gen_gauss_den_n5}.
%

%%%%%%%%%%%%%%%%%%%%%%%%%%%%%%%%%%%%%%%%%%%%%%%%%%%%%%%%%%
\section{DISCUSSION}\label{sec6}

Let us summarize the work done here. In this paper, we have investigated 
the gravitational collapse of matter fields in higher dimensional
general relativity. The gravitational field equations and equations governing the gravitational collapse
of matter fields depend on the dimensions of the spacetime. Using these equations, we verify 
that various features of gravitational collapse are indeed dependent on the dimensions of 
the spacetime. For example, as the spherical shell of matter collapses under gravity, one may 
easily determine the
following quantities, (a) the time of singularity formation, (b) the
time of horizon (MTT) formation as each shell collapses, (c)
the nature and signature of the horizon during the collapse process, and (d) the relative characteristics of
 the collapsing shells with respect to each other. Our studies in the previous sections show 
 that each of these features are dimension dependent. Therefore, 
 the time of singularity formation, in comparison to 
the horizon formation, is a dimension dependent statement. In addition to dimensionality, 
the dynamics of gravitational and matter fields also depend on the 
nature of matter model. We have discussed 
homogeneous and inhomogeneous dust models, as well as 
matter with anisotropic stress, and viscosity. In this variation of energy-momentum tensor fields undergoing collapse, we have also included deviations among them in terms of the initial density and velocity profiles. In each case, our matter models are realistic and should faithfully imitate
naturally occurring gravitational processes. In all of these studies, 
three distinct cases of importance have been carefully looked
into (i) the pressureless collapse scenario, (ii) matter models behaving like perfect fluids, and (iii) models including viscosity. These studies show us that (a) viscous fields hinder or accelerate the time of collapse of the gravitating matter and singularity formation time. However, for any such models, the singularity always remains covered by the horizon. Indeed, for all the choices of matter fields carried out by us, the density distribution is very close to realistic astrophysical scenarios, and so, we may safely conclude that our studies, which refer to a large class of realistic models, do support the censorship conjecture.

It however must be noted that the trapped surfaces studied in this paper are restrictive in 
the sense that they only cater to spherically symmetric foliations of the spacetime. One may certainly have nonspherical trapped surfaces, and these are much more abundant. Indeed, the trapped region is a collection of all such possible trapped surfaces and not just the spherical ones. Naturally, the boundary of this trapped region is the best candidate black hole horizon. However, no proof of this proposal exists, although, it is conjectured that the boundary of the outer trapped region must be the event horizon, and proof to that effect exists for a class of Vaidya spacetimes \cite{Eardley:1997hk,BenDov:2006vw}. In the present study, we have restricted to 
spherical cross sections and the spherical trapped surfaces only and considered its boundary as the candidate horizon. To be more inclusive, one has to include nonspherical trapped surfaces also which may also help us to include the matter models with nonspherical symmetry.

\section*{Acknowledgements}
The authors A. C. and A. K. are supported through the DAE-BRNS Project No. $58/14/25/2019$-BRNS.
A. C. acknowledges the support through the Department of Science and Technology scheme
of government of India through their Grant No. DST-MATRICS
MTR-/2019/000916.  
%The author AC and AK are supported through the DAE-BRNS project 58/14/25/2019-BRNS.
S. C. J. is thankful to the National Institute of Technology Hamirpur (NITH) and Central University of Jammu (CUJ) for providing necessary facilities during
the completion of this work.
A. K. is grateful to Ivan Booth for valuable discussion.
%%%%%%%%%%%%%%%%%%

\section{APPENDIX}\label{sec7}

\subsection{Derivation of the Einstein equations using tetrad basis}\label{sec_71}
The metric for $n$-dimensional spherical symmetric spacetime with spherical foliation is given by
(we shall use the gravitational units $G=1$, $c=1$ units in this paper)
\begin{equation}\label{ndmetric_app}
ds^2=-e^{2\alpha(r,t)}dt^2+e^{2\beta(r,t)}dr^2+R^2(r,t)\,d\Omega^2_{\, n-2}\,\,,
\end{equation}
where the spherical cross sections of dimension $(n-2)$ is given by the following form:
\begin{equation}\label{spheremetric_app}
d\Omega^2_{n-2}= \sum_{i=1}^{n-2} \left[ \prod_{j}^{i-1} \sin^{2} \,\theta^{\,j} \right](d\theta^{\,i})^{\,2}.
\end{equation}
Here, $\alpha(r,t)$, $\beta(r,t)$, and $R(r,t)$ are the three unknown functions to be determined. 
The variable $t$ denotes the time coordinate, $r$ denotes the radial coordinate, and the 
function $R(r,t)$ is radius of the spherical configuration of matter undergoing 
gravitational collapse.\\

In the following, we shall use the tetrad variables to obtain them. The tetrad basis $1$-forms suitable for the metric Eq. \eqref{ndmetric_app} are as follows:
\begin{equation}
e^{0}=e^{\alpha(r,t)}dt\, ,~~ e^{1}=e^{\beta(r,t)}\, dr\, ,~~ e^{i}=R(r,t)\, \theta^{i}\,,
\end{equation}
where $i=2,3,\dots,n$. The $(n-2)$ spherical metric is diagonal in this basis $d\Omega^2_{\, n-2}=\delta_{ij}\, \theta^{i}\,\theta^{j}$.  The spin connections can be obtained from 
the torsion-free condition, $de^I+\omega^{I}_{J}\wedge e^J=0$, where $I,J$ are the internal flat basis with 
metric $\eta_{IJ}$. The components of spin connection $\omega_{IJ}$ are obtained to be
\begin{eqnarray}
          {\omega^{0}}_{1}&=&\alpha^{\prime}e^{-\beta}\,e^{0}+\dot{\beta}e^{-\alpha}\,e^{1}\,\,,\hspace{1.5cm} {\omega^{0}}_{i}= (\dot{R}/R)\,e^{-\alpha}\,e^{i}\,\,,\\
          {\omega^{1}}_{i}&=&-(R^{\,\prime}/R)\,e^{-\beta}e^{i}\,, ~~~~~~~~~~~~~~~~~ {\omega^{i}}_{j}=\gamma^{i}{}_{j}\,\,,
\end{eqnarray}
where the spin connections are antisymmetric in their internal indices, $\omega_{IJ}=-\omega_{JI}$,
and the quantity $\gamma_{ij}$ is defined through the 
equation $d\theta^{\, i} =-\gamma^{i}{}_{j}\,\theta^{\,j}$. The primes indicate partial 
differentials with respect to the radial coordinate $r$ whereas dots indicate derivatives with respect 
to time. The curvature two form $\Omega_{IJ}$ are defined through
the equation $\Omega_{IJ}=d\omega_{IJ}+\omega_{IK}\wedge{\omega^{K}}_{J}$\,, and the nonzero components are the following:
\begin{eqnarray}
 {\Omega^{0}}_{1}&=&\left[\{e^{-\alpha}\,(e^{\beta})_{,t}\}_{,t}-\{e^{-\beta}\,(e^\alpha)_{,r}\}_{,r}    \right]e^{-(\alpha+\beta)} \, e^{0}\wedge e^{1}\,,\\
 {\Omega^{0}}_{i}&=&\left[\frac{e^{-\alpha}}{R}\,\{(R)_{,t}\,e^{-\alpha} \}_{,t}-\frac{R_{,r}}{R}\,(e^{\alpha})_{,r}\,e^{-(\alpha+2\beta)}\right]e^{0}\wedge e^{i}\nonumber\\
                      &&~~~~~~~~~~~~~~~~~~~~~~~~+ \left[\frac{e^{-\beta}}{R}\,\{(R)_{,t}\,e^{-\alpha} \}_{,r}-\frac{R_{,r}}{R}\,(e^{\beta})_{,t}\,e^{-(\alpha+2\beta)}\right]e^{1} \wedge e^{i}\, ,\\
          {\Omega^{1}}_{i}&=&\left[-\frac{e^{-\alpha}}{R}\,\{(R)_{,r}\,e^{-\beta} \}_{,t}+\frac{R_{,t}}{R}\,(e^{\alpha})_{,r}\,e^{-(2\alpha+\beta)}\right]e^{0}\wedge e^{i}\nonumber\\
                      &&~~~~~~~~~~~~~~~~~~~~~~~~+ \left[-\frac{e^{-\beta}}{R}\,\{(R)_{,r}\,e^{-\beta} \}_{,r}+\frac{R_{,t}}{R}\,(e^{\beta})_{,t}\,e^{-(2\alpha+\beta)}\right]e^{1} \wedge e^{i} \, ,\\
          {\Omega^{i}}_{j}&=&\left[\left(\frac{R_{,t}}{e^{\alpha}R} \right)^2-\left(\frac{R_{,r}}{e^{\beta}R} \right)^2\right]e^i \wedge e^j + \mathcal{R}^{i}{}_{j} \, ,
\end{eqnarray} 
where $\mathcal{R}_{ij}$ is the curvature $2-$form of the $(n-2) $ dimensional submanifold, and is given by $\mathcal{R}_{ij}=d\omega_{ij}+\omega_{ik} \wedge \omega^{k}{}_{j}$. The components of the curvature two forms can be extracted by using $\Omega_{IJ}=(1/2)\Omega_{IJKL}e^{K}\wedge e^{L}$, and the Ricci tensor components are given by $R_{IJ}={\Omega^{K}}_{IKJ}$. The expressions for these quantities are as follows:
\begin{eqnarray}
R_{00}&=&\Omega^{1}{}_{010}+\sum_{i}\, \Omega^{0}{}_{i0i}= -e^{-(\alpha+\beta)}\left[\{e^{-\alpha}(e^{\beta})_{,t}\}_{,t}\right]+e^{-(\alpha+\beta)}\left[\{e^{-\beta}(e^{\alpha})_{,r}\}_{,r}\right]\\
&& ~~~~~~~~~~~~~~~~~~~~~~~ - (n-2)\left[(e^{-\alpha}/R)
\{\dot{R}e^{-\alpha}\}_{,t}-e^{-(\alpha+2\beta)} (R^{\prime}/R)\, \{e^{\alpha}\}_{,r}\right],\nonumber\\
R_{01}&=&-\sum_{i}\Omega^{0}{}_{i1i}=-(n-2)\left[(e^{-\beta}/R)\{\dot{R}e^{-\alpha}\}_{,r}
-e^{-(\alpha+2\beta)}(R^{\prime}/R)(e^{\beta})_{,t}\right],\\
R_{\,0\,i}&=&R_{1 i}=0,\\
R_{ij}&=&\delta_{ij}\left[\frac{e^{-\alpha}}{R}(e^{-\alpha}\dot{R})_{,t}-
\frac{e^{-(2\beta+\alpha)}}{R}\,R^{\prime}(\,e^{\alpha})_{,r}-\frac{e^{-\beta}}{R}(e^{-\beta} R^{\,\prime})_{,r}+\frac{e^{-(2\alpha+\beta)}}{R}\dot{R}(e^{\beta})_{,t}\right]\nonumber\\
&&~~~\hfill+(n-3)\,\delta_{ij}\left[\left(\frac{\dot{R}e^{-\alpha}}{R}\right)^{2}-\left(\frac{{R}^{\,\prime}
e^{-\beta}}{R}\right)^{2}+\frac{\eta}{R^{2}}\right],
\end{eqnarray}
where $\eta=0,- 1. +1$ for planer, hyperbolic and spherical geometries respectively. For the present case, we shall use $\eta=1$.  The Ricci scalar for this metric is $R=\eta_{IJ}R^{IJ}$ and gives
\begin{eqnarray}
R&=&-R_{00}+R_{11}+\sum_{i}R_{ii}\nonumber\\
&=&2e^{-(\alpha+\beta)}\left[\frac{\partial}{\partial t}\{e^{-\alpha}(e^{\beta})_{\,,t}\}-\frac{\partial}{\partial r}\{e^{-\beta}(e^{\alpha})_{\,,r}\} \right]+(n-2)(n-3)\left[\frac{e^{-2\alpha}\dot{R}^{2}}{R^{2}}- \frac{e^{-2\beta}{R}^{\,\prime \,2}}{R^{2}}+\frac{\eta}{R^{2}}\right]\nonumber\\
&+&2(n-2)\left[\frac{e^{-\alpha}}{R}\frac{\partial}{\partial t}(e^{-\alpha}\dot{R})-\frac{e^{-(2\beta+\alpha)}}{R}\frac{\partial}{\partial t}(Re^{\alpha})-\frac{e^{-\beta}}{R}\frac{\partial}{\partial r}
(R^{\prime}e^{-\beta})+\frac{e^{-(\beta+2\alpha)}}{R}\frac{\partial}{\partial t}(Re^{\beta})\right].
\end{eqnarray}
The Einstein tensor in the orthonormal frame is easily obtained by using equation $G_{IJ}=R_{IJ}-(1/2)\eta_{IJ}R$. However, we only need the $G_{00}$, $G_{01}$ and the $G_{11}$ equations, while the $G_{ij}$
equations shall not be used, but shall instead be replaced by the Bianchi identities.
\begin{eqnarray}
G_{00}&=&(n-2)\left[-\frac{e^{-\beta}}{R}\frac{\partial}{\partial r}(e^{-\beta}R^{\prime})
          +\frac{e^{-(2\alpha+\beta)}}{R}\dot{R}\frac{\partial}{\partial t}(e^{\beta})\right] +H , \label{g00_eqn_app}\\
G_{11}&=&(n-2)\left[-\frac{e^{-\alpha}}{R}\frac{\partial}{\partial t}(e^{-\alpha}\dot{R})
          +\frac{e^{-(\alpha+2\beta)}}{R}R^{\prime}\frac{\partial}{\partial r}(e^{\alpha})\right] -H ,\label{g11_eqn_app}\\
H&=&\frac{(n-2)(n-3)}{2}\left[\frac{e^{-2\alpha}}{R^{2}}\, \dot{R}^{2}-\frac{e^{-2\beta}}{R^{2}}\, R^{\,\prime \,2} +\frac{\eta}{R^{2}}\right],\\
G_{01}&=& \frac{(n-2)}{R}e^{-(\alpha+\beta)}[\dot{R}^{\,\prime}-\alpha^{\prime}\dot{R}-\dot{\beta}R^{\,\prime}]\label{r01_eqn_app}.
\end{eqnarray}
To find the Binachi identities,
let us now have a look at the Bianchi identities, $\nabla_{\mu}T^{\mu\nu}=0$.
The equation for time component is
\begin{eqnarray}
\dot{\rho}e^{-\alpha(r,t)}+\theta(\rho +p_{t})+(p_{r}-p_{t})\dot{\beta}e^{-\alpha}
-\frac{2(n-2)}{(n-1)}\,\eta\sigma^{2}-\zeta\theta^{2}=0,
\end{eqnarray} 
which on rearranging, may be put in the following expression for $\dot{\beta}$:
\begin{eqnarray}\label{betadot_eqn}
\dot{\beta}&=&\frac{-\dot{\rho}}{\left[\rho+p_{r}-\frac{2(n-2)}{(n-1)}\,\eta\sigma-\zeta\theta \right]}-(n-2)\,\frac{\dot{R}}{R}\,\frac{\left[ \rho+p_{t}+\frac{2}{(n-1)}\,\eta\sigma-\zeta\theta \right]}{\left[\rho+p_{r}-\frac{2(n-2)}{(n-1)}\,\eta\sigma-\zeta\theta \right]}\,\,.
\end{eqnarray}
The $r$-component is similarly obtained to be
\begin{eqnarray}
&&(p_{t}-\zeta\theta)^{\,\prime}+(p_{r}-p_{t})^{\, \prime}+\alpha^{\prime}\,(\rho +p_{r}-\zeta\,\theta)
+(n-2)(p_{r}-p_{t})\frac{R^{\,\prime}}{R}\nonumber\\
&&~~~~~~~~~~~~~~~~~~~~~~-\frac{2(n-2)}{(n-1)}\eta\sigma^{\,\prime}
-\frac{2(n-2)}{(n-1)}\eta\sigma\,\left[\alpha^{\prime}+(n-1)\frac{R^{\,\prime}}{R}\right]=0.
\end{eqnarray}
This gives us the expression for the quantity $\alpha^{\prime}$ in terms of the other metric and matter variables:
\begin{eqnarray}\label{alphaprime_eqn_app}
\alpha^{\,\prime}&=& -(n-2)\frac{R^{\,\prime}}{R}\frac{\left(p_{r}-p_{t}-2\eta\sigma \right)}{\left[\rho+p_{r}-\frac{2(n-2)}{(n-1)}\,\eta\sigma-\zeta\theta \right]}
-\frac{\left[p_{r}-\frac{2(n-2)}{(n-1)}\,\eta\sigma-\zeta\theta \right]'}{\left[\rho+p_{r}-\frac{2(n-2)}{(n-1)}\,\eta\sigma-\zeta\theta \right]}\,.
\end{eqnarray}
The $R_{01}$ component of the Einstein equation is given by Eq. \eqref{r01_eqn_app}. Using Eqs. \eqref{betadot_eqn} and \eqref{alphaprime_eqn_app} in Eq. \eqref{r01_eqn_app} and multiplying $R^{n-2}$, we get
\begin{equation}
\left[\left\{p_{r}-\frac{2(n-2)}{n-1}\,\eta\sigma-\zeta\theta \right\}\frac{2R^{n-2}\dot{R}}{(n-2)} \right]_{,r}
+\left[\frac{2\rho R^{n-2}R'}{(n-2)} \right]_{,t}=0\,.
\end{equation}
These equations may now be interpreted as the differential of a function $F(r,t)$ as follows:
\begin{eqnarray}\label{frt_guess_app}
F^{\prime}(r,t)&\propto & \rho R^{n-2}R^{\prime}\,\,,\\
\dot{F}(r,t)&\propto & \left[p_{r}-\frac{2(n-2)\eta\sigma}{n-1}-\zeta\theta \right]R^{n-2}\dot{R} \,.
\end{eqnarray}
This function $F(r,t)$ shall be called the (Misner- Sharp) mass function. 
To determine the exact form of $F(r,t)$, we multiply $G_{00}$, Eq. \eqref{g00_eqn_app}, with 
$R^{\prime}\, R^{(n-3)}$ and $G_{11}$, Eq. \eqref{g11_eqn_app}, with $\dot{R}\,R^{(n-3)}$ :
\begin{eqnarray}\label{frt_form}
\left[ R^{(n-3)}\left(1+{\dot{R}}^{2}e^{-2\alpha}-{R}^{\prime\,2}e^{-2\beta} \right)\right]_{\,,r}&=&\frac{2}{(n-2)}\,\rho R^{n-2}R^{\prime}\,,\\
\left[ R^{(n-3)}\left(1+{\dot{R}}^{2}e^{-2\alpha}-{R}^{\prime\, 2}e^{-2\beta} \right)\right]_{\,,t}&=&
-\frac{2}{n-2}\left[p_{r}-\frac{2(n-2)}{(n-1)}\eta\sigma-\zeta\theta \right] \dot{R}R^{\,n-2}\,. 
\end{eqnarray}
Comparing Eqs. \eqref{frt_guess_app} and \eqref{frt_form}, the mass function is giving by following form:
\begin{equation}\label{massfunction_app}
F(r,t)= R^{\,(n-3)}\,(r,t)\left[1+{\dot{R}^{\,2}\,(r,t)}e^{-2\alpha}-{R}^{\prime\,2}\,(r,t)\,e^{-2\beta} \right].
\end{equation}
Defining the two functions, $H(r,t)=e^{-2\alpha(r,t)}\,{\dot{R}}^{2}$ and $G(r,t)=e^{-2\beta(r,t)}\,R^{\prime\,2}$, the Einstein equations reduce to the following set:
 \begin{eqnarray}\label{grav_coll_eqn_appendix_app}
 \rho(r,t)&=&\frac{(n-2)}{2R^{n-2}R'}\,F^{\prime}\,(r,t),\\%\hspace{2cm}
 p_{r}(r,t)&=&\frac{2(n-2)}{(n-1)}\eta\sigma+\zeta\theta-\frac{(n-2)\dot{F}}{2R^{n-2}\dot{R}} \,\,,\\
 \alpha(r,t)^{\prime}&=& -\frac{(n-2)R'}{R}\frac{\left(p_{r}-p_{t}-2\eta\sigma \right)}{\left[\rho+p_{r}-\frac{2(n-2)}{(n-1)}\eta\sigma-\zeta\theta \right]}-\frac{\left[p_{r}-\frac{2(n-2)}{(n-1)}\eta\sigma-\zeta\theta \right]'}{\left[\rho+p_{r}-\frac{2(n-2)}{(n-1)}\eta\sigma-\zeta\theta \right]}  ,\\
 2{\dot{R}}'&=&R'\frac{\dot{G}}{G}+\dot{R}\frac{H'}{H}\,\,,\\
F(r,t)&=&R^{(n-3)}\left(1-G+H \right)\,.
\end{eqnarray}
These are the complete and independent set of Einstein equations for the study of spherical gravitational collapse of viscous fluid. These have been used in the main text.\\

%%%%%%%%%%%%%%%%%%%%%%%%%%%%%%%%%%%%%%%%%%%%%%%
\subsection{Newman- Penrose scalar $\psi_{2}$ in higher dimensions}\label{sec72}
In this paper, we have made use of the Newman- Penrose (NP) formalism while calculating many of the 
functions residing on the horizon. In this section, we collect some of the basic equations that
are useful in the present context.  Firstly, we denote the $n$-dimensional spacetime metric by
\begin{equation}
g_{ab}=-2\ell_{(a}\, n_{b)}+\delta_{ij}\, m^{i}_{a}m^{j}_{b},
\end{equation}
where the indices $a, b, \cdots$ take values from $0$ to $n-1$, while $i, j,\cdots$ are the indices on the submanifold, and take values from $2$ to $n-1$. The lowercase indices $i, j \cdots$ shall be raised and lowered using $\delta_{ij}$. The vectors $\ell^{a}$, and $n^{a}$ are null while 
the $m^{i}_{a}$\,s are orthonormal and spacelike. These vectors have the following inner products:
\begin{equation}
\ell^{a}\ell_{a}=0= n^{a}\, n_{a}=\ell^{a}\, m^{i}_{a}=n^{a}m^{i}_{a}, ~~~ \ell^{a}n_{a}=-1, ~~~m^{i\,a}\, m^{j}_{a}=\delta_{ij}. 
\end{equation}
The covariant derivatives of the vectors $(\ell, n, m^{i}_{a})$ are called the Ricci coefficients and are defined as follows:
\begin{equation}
\nabla_{b}\,\ell_{a}=L_{cd}\, m^{c}_{a}\, m^{d}_{b}, ~~ \nabla_{b}\,n_{a}=N_{cd}\, m^{c}_{a}\, m^{d}_{b}, ~~\nabla_{b}\,m^{i}_{a}=M^{i}_{cd}\, m^{c}_{a}\, m^{d}_{b},
\end{equation}
which satisfy the constraints 
\begin{eqnarray}
&& L_{0a}=N_{1a}=0,~~N_{0a}+L_{1a}=0, \nonumber\\
&& M^{i}_{0a}+L_{ia}=0, ~~M^{i}_{1a}+N_{ia}=0, ~~M^{i}_{ja}+M^{j}_{ia}=0.
\end{eqnarray}
These Ricci coefficients combine to form 
the NP scalars. In $4D$, various components of $L$ add to form 
the NP $\kappa, \rho, \sigma$, and so on. The full details of 
these quantities in $4D$ may be found in \cite{Chandrasekhar}, and for
arbitrary dimensions, in \cite{Coley:2004jv,Ortaggio:2007eg}.   
In our calculations, we have used the decomposition of the Weyl tensor
$C_{abcd}$, in terms of its \emph{boost weight} see \cite{Coley:2004jv,Ortaggio:2007eg}.
For the boost weight zero components of the Weyl tensor, the quantities
$C_{0101}$ and $C_{ijij}$ are important for the computation of $\psi_{2}$,
where these are defined through
\begin{equation}
C_{abcd}=C_{ijij}\, m^{i}_{\{ a}m^{j}_{b}m^{i}_{c}m^{j}_{d\}}\,\, , ~~~C_{abcd}=C_{0101}\, \ell_{\{ a}n_{b}n_{c}\ell_{d\}}. 
\end{equation}
These two quantities are related due to the symmetry properties of the Weyl tensor:
$C_{0101}=(-1/2)C_{ijij}$. Thus, instead of determining the $\psi_{2}$ using $C_{ijij}$, we use the quantity $C_{0101}$.

 %%%%%%%%%%%%%%%%%%%%%%%%%%%%%%%%
 %%%%%%%%%%%%%%%%%%%%%%%%%%%%%%
 %%%%%%%%%%%%%%%%%%%%%%%%%%%
 \subsection{Geometrical equations on the spatial subspace}\label{sec73}
\label{geometry_appendix}
Many of the equations used in the main text are fundamental equations on submanifolds. 
The double null frame used in our problem implies that the subspace is of codimension 2. In 
the following, we shall briefly compile the main equations for such a subspace and present a brief 
outline of their derivations.
 
Let $(\mathcal{M},\, g_{\mu\nu}, \nabla_{\mu})$ be a $n$-dimensional, orientable, 
time-orientable spacetime with a metric compatible covariant 
derivative $\nabla$, such that $\nabla_{\mu}\,g_{\nu\lambda}=0$.  $\mathcal{S}$ is a closed, orientable, spacelike $(n-2)$ surface in $\mathcal{M}$.
Let $\ell^{\mu}$ and
$n^{\mu}$, be the outgoing and incoming null vector fields such that $\ell\cdot n=-1$. 
The induced metric $h_{ab}$ on 
the subspace $\mathcal{S}$ is given by
\begin{equation}
h_{ab}=e^{\mu}{}_{a}\, e^{\nu}{}_{b}\, g_{\mu \nu},
\end{equation}
where $a, b, \dots$ indicate indices on $\mathcal{S}$, and $e^{\mu}{}_{a}$ is the pullback tensor and 
is orthogonal to  both $\ell^{\mu}$ and $n^{\mu}$. The inverse 
of metric of $(n-2)$ dimensional subspace $h^{ab}$ is given by
\begin{equation}
g^{\mu\nu}=e^{\mu}{}_{a}e^{\nu}{}_{b}\, h^{ab}- \ell^{\,\mu}n^{\nu}- \ell^{\,\nu}n^{\mu}.
\end{equation}
Corresponding to each of the null normals, we get one extrinsic curvature or the second fundamental forms. The extrinsic curvature is a vector on the normal bundle $N(\mathcal{S})$ of $\mathcal{S}$ and is given by
\begin{equation}
k^{\,\mu}{}_{ab}=k^{(n)}{}_{ab}\,\ell^{\mu}+k^{(\ell)}{}_{ab}\, n^{\mu}.
\end{equation}
The two components of the extrinsic curvature, along $\ell^{\mu}$ and $n^{\mu}$ are given by
\begin{equation}
k^{(\ell)}{}_{ab}=e^{\mu}{}_{a}e^{\nu}{}_{b}\, \nabla_{\mu}\, \ell_{\nu}, ~~~~~~
k^{(n)}{}_{ab}=e^{\mu}{}_{a}e^{\nu}{}_{b}\, \nabla_{\mu}\,n_{\nu} \,.
\end{equation}
The Riemann tensor on $\mathcal{M}$ and on $\mathcal{S}$ are given, respectively, by
\begin{eqnarray}
(\nabla_{\mu}\nabla_{\nu}- \nabla_{\nu}\nabla_{\mu})Z_{\lambda}&=&R_{\mu\nu\lambda\sigma}\,Z^{\sigma}\\
(\mcD_{a}\mcD_{b}- \mcD_{b}\mcD_{a})z_{c}&=&\mathcal{R}_{abcd}\,z^{d},
\end{eqnarray}
where $\mcD$ is the metric compatible derivative operator on $\mcS$, so that $\mcD_{a}h_{bc}=0$.
The two fundamental equations of submanifolds are the Gauss and the Codazzi equations. The Gauss equation relates the pullback of the (full spacetime) Riemann tensors to the subspace Riemann tensor defined above. This relation is given by the following equation:
\begin{equation}
e^{\mu}{}_{a}e^{\nu}{}_{b}e^{\lambda}{}_{c}e^{\sigma}{}_{d}\, R_{\mu\nu\lambda\sigma}
=\mcR_{abcd}-\{k^{(\ell)}{}_{ac}\,k^{(n)}{}_{bd}+k^{(n)}{}_{ac}\,k^{(\ell)}{}_{bd}\}+
\{k^{(\ell)}{}_{ad}\,k^{(n)}{}_{bc}+k^{(n)}{}_{ad}\, k^{(\ell)}{}_{bc}\},
\end{equation}
The Codazzi equations are projections of the (pullback) Riemann tensor along each of the two null normals $\ell^{\mu}$, and $n^{\mu}$. The corresponding equations are
\begin{eqnarray}
e^{\mu}{}_{a}e^{\nu}{}_{b}e^{\lambda}{}_{c}\,\ell^{\sigma}\, R_{\mu\nu\lambda\sigma}
&=&(D_{b}-\omega_{b})k^{(\ell)}{}_{ac} - (D_{a}-\omega_{a})k^{(\ell)}{}_{bc}\\
e^{\mu}{}_{a}e^{\nu}{}_{b}e^{\lambda}{}_{c}\, n^{\sigma}\, R_{\mu\nu\lambda\sigma}
&=&(D_{b}-\omega_{b})k^{(n)}{}_{ac} - (D_{a}-\omega_{a})k^{(n)}{}_{bc}.
\end{eqnarray}
In the above equations, the $1$- form $\omega$ is related to the shape operator,
$\omega_{\underleftarrow{a}}=-n_{\sigma}\,e^{\lambda}{}_{a}\,\nabla_{\lambda}\ell^{\sigma}$.
Note also that $\omega_{\underleftarrow{a}}\equiv e^{\mu}{}_{a}\,\omega_{\mu}$ is 
the pullback of full spacetime connection on the normal bundle $\mathcal{N}(\mcS)$.

The derivative of geometric quantities defined on the submanifold $\mcS$, (like $h_{ab}$ and $k_{ab}$) in 
the direction normal to it are useful for simplification of the above equations. Let 
us define a normal by $N^{\mu}=A\ell^{\mu}-Bn^{\mu}$ where $A$ and $B$ are functions on foliation.
For example, the derivative of the induced metric is
\begin{equation}
\nabla_{N}h_{ab}=2Ak^{(\ell)}_{ab}-2Bk^{(n)}_{ab}\, .
\end{equation}
Using this, the variation of area element $\sqrt{h}=\sqrt{\det h_{ab} }$ is given by
\begin{equation}
\nabla_{N}\sqrt{h}=(1/2)\sqrt{h}\, h^{ab}\,\nabla_{N}h_{ab}=[\, A\theta_{(\ell)}-B\theta_{(n)}\,]\sqrt{h}\, ,
\end{equation}
where $\theta_{(\ell)}$ and $\theta_{(n)}$ are the expansion scalars of the null congruences
corresponding to the null directions $\ell^{a}$ and $n^{a}$. This relation may be derived from the 
relation of extrinsic curvature with expansion scalars and the shear tensors:
\begin{equation}
k^{(\ell)}{}_{ab}=\frac{1}{(n-2)}\,\theta_{(\ell)}h_{ab}+\sigma^{(\ell)}_{(ab)},~~~~
k^{(n)}{}_{ab}=\frac{1}{(n-2)}\, \theta_{(n)}h_{ab}+\sigma^{(n)}_{(ab)} 
\end{equation}
where the expansion scalar and the shear tensors are defined as:
\begin{eqnarray}
\theta_{(\ell)}&=&\nabla_{\mu}\ell^{\mu}-\kappa_{(\ell)}\\
\sigma^{(\ell)}_{ab}&=&\left[e^{\mu}{}_{a}e^{\nu}{}_{b}-
\frac{h_{ab}}{(n-2)}\, g^{\mu\nu}\right]\nabla_{\mu}\ell_{\nu}
+\kappa_{(\ell)}\,h_{ab},
\end{eqnarray}
where $\kappa_{(\ell)}=-n_{\nu}\ell^{\mu}\,\nabla_{\mu}\ell^{\,\nu}$.
The equations for $n^{\mu}$
is obtained by substituting $\ell^{\mu}\leftrightarrow n^{\mu}$ in the above equations. \\

These equations are valid on a given foliation of the submanifold. As time evolves, the foliation and the geometry evolves with it. Let us determine the evolution of foliation along $N^{\mu}$. 
First, the pullback of null normals $\ell_{\mu}$ and $n_{\mu}$
on $\mcS$ must vanish by definition: $e^{\mu}{}_{a}\ell_{\mu}\equiv \ell_{\underleftarrow{a}}=0$,
and $e^{\mu}{}_{a}n_{\mu}\equiv n_{\underleftarrow{a}}=0$. 
Second, we shall assume that foliation be preserved under 
$N^{\mu}$. This implies that $(\lie_{N}\,\ell)_{\underleftarrow{a}}=0$, and 
$(\lie_{N}\,n)_{\underleftarrow{a}}=0$. These equations lead to
\begin{eqnarray}
N^{\mu}\nabla_{\mu}\ell_{\underleftarrow{a}}&=&\kappa_{(N)}\ell_{\underleftarrow{a}}
-(D_{\underleftarrow{a}}-\omega_{\underleftarrow{a}})B,\\
N^{\mu}\nabla_{\mu}n_{\underleftarrow{a}}&=&-\kappa_{(N)}n_{\underleftarrow{a}}
+(D_{\underleftarrow{a}}+\omega_{\underleftarrow{a}})B,
\end{eqnarray}
where $\kappa_{(N)}=-n_{\mu}N^{\nu}\nabla_{\nu}\ell^{\mu}$ is called the surface gravity
corresponding to the vector field $N^{\mu}$.\\

 A direct calculation leads to the following results on 
the variation of $\theta_{(\ell)}$ \cite{Booth:2006bn}
\begin{eqnarray}\label{deriv_eqn}
\nabla_{N}\,\theta_{(\ell)}-\kappa_{N}\,\theta_{(\ell)}&=&-d^{2}B+2\omega^{\mu}d_{\mu}B- B\,[\,|\,\omega \,|^{2}
-d\omega
-(\mathcal{R}/2)+G_{\mu\nu}\,\ell^{\mu}\,n^{\nu}-\theta_{(\ell)}\,\theta_{(n)}\nonumber]\\
&& ~~~~~~~~~~~~~~~~~~~~~~~~~~~~~~ -A[\,\sigma_{(\ell)}^{2}+G_{\mu\nu}\,\ell^{\mu}\,\ell^{\nu} +(1/2)\,\theta_{(\ell)}^{2}\,].
\end{eqnarray}
Here, $\mathcal{R}$ is the scalar curvature of the $(n-2)$ spacelike sphere. In the present work, we are using the standard $t$=constant, $r$= constant foliations: The $\ell^{\mu}$ and $n^{\mu}$
are orthogonal to these foliations. As a result, the normal $N^{\mu}$ has no dependence on the angular coordinates of the metric Eq. \eqref{ndmetric}. To preserve this particular foliation, the diffeomorphisms 
generated by $N^{\mu}$ should map round spheres to round spheres. Therefore, the normal vector $N^{\mu}=A\, \ell^{\mu}- Bn^{\mu}$ should be such that $A,\, B$ are independent of all the angular coordinates. To be precise, they shall be taken to be constants. 

To find the parameter $C$ in Eq. \eqref{c_eqn}, let 
us choose (a) $A=1$, and $B=0$
in Eq. \eqref{deriv_eqn}.
On the spherical MTT, where $\theta_{(\ell)}=0$, and shear vanishes, this gives
\begin{eqnarray}\label{deriv_eqn1}
\lie_{\ell}\theta_{(\ell)}=-G_{\mu\nu}\ell^{\mu}\ell^{\nu} ,
\end{eqnarray}
whereas, on using (b) $A=0$, and $B=-1$, the same equation gives
\begin{eqnarray}\label{deriv_eqn2}
\lie_{n}\theta_{(\ell)}= -\mathcal{R}/2 +G_{\mu\nu}\ell^{\mu}n^{\nu} .
\end{eqnarray}
These expressions in Eqs. \eqref{deriv_eqn1}, and \eqref{deriv_eqn2}, along with the Einstein
equations lead directly to 
the value of $C$ given by \eqref{c_eqn}.
Using Einstein equations and Eq. \eqref{c_eqn}, the equation for $C$ is given by
\begin{eqnarray}
C=\frac{T_{ab}l^{a}l^{b}}{\mathcal{R}/2-T_{ab}l^{a}n^{b}}.
\end{eqnarray}
Now, for a $(n-2)$ dimensional round sphere of radius $R$, its area $\mathcal{A}_{n-2}$ is given by
\begin{equation}
\mathcal{A}_{n-2}=\frac{2\pi^{(n-1)/2}}{\Gamma({\frac{n-1}{2}})}\,R^{(n-2)}.
\end{equation}
By substituting this value of area in terms of the Ricci scalar of $(n-2)$- dimensional Ricci scalar 
$\mathcal{R}=(n-2)(n-3)/R^{2}$, and the radius of the sphere, we get a simpler expression.

For the collapse of pressureless dust cloud, and the expression of the null normals:
\begin{eqnarray}
    l^{a}&=&(\partial t)^{a}+ e^{-\beta(r,t)}(\partial r)^{a}\\
    n^{a}&=&(1/2)(\partial t)^{a}-(1/2)\, e^{-\beta(r,t)}(\partial r)^{a},
\end{eqnarray}
the expression for $C$ reduces to the following expression:
\begin{equation}
      C=\frac{2\,F(r)'}{2\,(n-3)\,R(r,t)^{\prime}\,R_{2m}^{(n-4)}-F(r)'}\,. 
\end{equation}
The expression for $C$ in case of viscous fluids is modified in case the mass function is 
independent of time to the following value:
\begin{equation}
  C=\frac{(1/2)\,F(r)'}{2\,(n-3)\,R(r,t)'\,R_{2m}^{(n-4)}-F(r)'}\, ,  
\end{equation}
whereas, for a time dependent mass function, the expression above is further modified:
\begin{equation}
    C=\frac{1}{2}\left[\frac{\,F'}{2\,(n-3)\,R'\,R_{2m}^{(n-4)}-F'}+\frac{(p_{r}-\bar{\eta}\sigma-\zeta\theta)}{(n-2)(n-3)R_{2m}^{-2}+(p_{r}-\bar{\eta}\sigma-\zeta\theta)}\right]\, .
\end{equation}
These expressions have been used in the main part of the paper to calculate 
the value of $C$ and hence, determine the stability of the corresponding MTTs.

%%%%%%%%%%%%%%%
\subsection{Bound collapse of matter configurations} \label{bound_coll}     
For $k>0$, to relate the shell radius $R(r,t)$ with time $t$, we need to solve 
the equation Eq. \eqref{shelleqn_gen}:
\begin{equation}\label{tr_eqn_hb}
dt=-\frac{dR}{\sqrt{\frac{F(r)}{R^{2(N-1)}} -k}}.
\end{equation}
To solve this equation, we use the  following parametrisation
for the shell radius $R(\eta,r)$:
\begin{eqnarray}\label{shell_hb}
R(\eta,r)&=&\left[\frac{F(r)}{k(r)}\,\cos^{2}\left(\eta/2 \right) \right]^{\frac{1}{2(N-1)}}\,\, ,\end{eqnarray}
where, the new time coordinate $\eta$ is related to the old time coordinate $t$ through the relation:
\begin{equation}
d\eta=\frac{2(N-1)\sqrt{k}}{R}\, dt
\end{equation}
Using this coordinate transformation in Eq. \eqref{shelleqn_gen}, we get
\begin{equation}
\frac{dR}{d\eta}=-\frac{R}{2(N-1)}\left[-1+\frac{F(r)}{k\,R^{2(N-1)}}\right]^{1/2}.
\end{equation}       
Equation \eqref{tr_eqn_hb} may be integrated by using Eq. \eqref{shell_hb}, giving the time curve for 
the shell marked $r$, in terms of the proper time $\eta$:
\begin{eqnarray}\label{time_hb}
 t(\eta,r)&=&-\frac{1}{N}\left[F(r)\,\,k(r)^{(N-2)}\right]^{\frac{1}{2(N-1)}}\,\cos\left(\eta/2 \right)^\frac{N}{(N-1)}\,\,{{}_{2}}F_{1}\left[ \frac{1}{2};\,\frac{N}{2(N-1)};\,\frac{3N-2}{2(N-1)};\,\cos^{2}\left(\eta/2\right)\right]\nonumber\\
            &&+\,\,\frac{\,\sqrt{\pi}\,\,\Gamma\left[ (3N-2)/2(N-1)\right]}{N\,\Gamma\left[(2N-1)/2(N-1)\right]}\,F(r)^{1/2(N-1)}\,\,\, k(r) ^{\frac{N-2}{2(N-1)}}\,\,.
\end{eqnarray}
In Eqn. \eqref{shell_hb} and \eqref{time_hb},  using $k(r)=r^{2}$ and $F(r)=m\,r^{2N}$, 
the expression for radius and time coordinates 
shows that collapse begins at $\eta=0$, and the starting time $t_{i}=0$, and it reaches 
the singularity $R=0$ at $\eta=\pi$. To simplify the calculations, we rewrite the equation Eq. \eqref{shell_hb} as
\begin{equation}
{R(\eta,r)=\left[(R_{0}/2) \,\left(1+\cos\eta \right) \right]^{\frac{1}{2(N-1)}}}.
\end{equation}
The time taken by 
the shell initially at $R_{0}=m\,r^{2(N-1)}$ to reach its Schwarzschild radius at $R=\left(2M\right)^{\frac{1}{2(N-1)}}$ is given by
\begin{equation}
\eta_{2M}=\cos^{-1}\,\left[\frac{4 M}{R_{0}}-1 \right].
 \end{equation}
The above equation represents the time of formation of MTT for each collapsing shell of mass $M$.
This junction matching give us 
the following conditions :
\begin{equation}
R(t)=a(t)\sin\chi\,; \hspace{2cm} 2M=F(r).
\end{equation}
The shell collapse begins at $\eta=0$, which implies that
\begin{equation}
R(t=0)\equiv R_{0}=a_{0}\sin\chi_{0}\,; \hspace{1.5cm} 2M=F(r_{0})\,,
\end{equation}
where $r_{0}=\sin\chi_{0}$ is the initial radial coordinate of the shell. From these two equations,
we conclude that
 \begin{equation}
\sin\chi_{0}=\left[\frac{2M}{R_{0}} \right]^{\frac{1}{2(N-1)}}\,; \hspace{1.5cm} a_{0}^{(2N-1)}=\frac{R_{0}^{2N}}{\sqrt{2M}}\,.
 \end{equation}
The time taken for the matter shells to arrive at the singularity $R=0$ and $\eta=\pi$ is given by
 \begin{equation}\label{sing_time_hb}
 t_{s}(r)=\frac{\,\sqrt{\pi}\,\Gamma\left[ \frac{3}{2}+\frac{1}{2(N-1)}\right]}{N\,\Gamma\left[1+\frac{1}{2(N-1)}\right]}\,\left[\frac{F(r)}{k(r)} \right]^{\frac{1}{2(N-1)}}\sqrt{k(r)}\,.
 \end{equation}
 Since $F=m\,r^{2N}$, and here $m$ is constant, Eq. \eqref{sing_time_hb}, implies that
 the shells with different initial radii reach the central singularity at the same time. The outermost trapped surface is identified through the equation $R(r,t)=F(r)^{\frac{1}{2(N-1)}}$.
 From Eq. \eqref{shell_hb}, we get 
 \begin{equation}
 \eta=2\, \cos^{-1}\left[\sqrt{\frac{k}{F(r)}}\,\,R(r,t)^{(N-1)} \right]
\end{equation} 
 and hence, the MTT is described by
 \begin{equation}
\eta_{_{\scaleto{MTT}{4pt}}}=2\cos^{-1}r=2\cos^{-1}(\sin\chi)=\pi-2\chi.
\end{equation}
\\
To study the formation and evolution of the event horizon, we shall need 
the null geodesics which, in the $(r,t)$ coordinates, are given by,
 \begin{equation}
 \frac{dr}{dt}=\frac{\sqrt{1-k(r)}}{R'}=\frac{\sqrt{1-r^{2}}}{\left[m\cos^{2}\left(\eta/2 \right) \right]^{\frac{1}{2(N-1)}}}\,.
 \end{equation}
 Now use the condition that the last shell and the last outgoing null geodesic 
 reach the horizon point at same instant.
 So, for this shell, the boundary condition is $\chi=\chi_{0}$ at $\eta=\eta_{2M}=\pi -2\chi_{0}$, and the coordinates of the event horizon are given by
\begin{equation}
\chi_{eh}=\chi_{0}+(\eta-\eta_{2M})\,.
\end{equation} 
From the previous equation and Eq. \eqref{shell_hb}, the last outgoing null ray inside the cloud is given by
\begin{equation}
 R_{eh}=r_{eh}\left[m\cos^{2}\left(\eta/2 \right) \right]^{\frac{1}{2(N-1)}}=\sin\left(\chi_{0}+\eta-\eta_{2M} \right)\left[m\cos^{2}\left(\eta/2 \right) \right]^{\frac{1}{2(N-1)}}\,.
\end{equation}
\subsection{Junction conditions}\label{sec75}
In the following, we shall present general conditions for matching two metrics 
across a junction. The interior spacetime $\it\bf(M^-)$ is given by the equation:
\begin{equation}
ds^{2}=-e^{2\alpha(r,t)}dt^2 + e^{2\beta(r,t)}dr^2 + R(r,t)^2 d \Omega^2_{n-2}.
\label{m1}
\end{equation}
The line element of the exterior spacetime is taken to be a generalisation of the Vaidya spacetime:
\begin{equation}
ds^{2}=-\left[1-\frac{2M(v)}{r^{2(N-1)}}\right]dv^2 -2 dv\, dr + r^2      d \Omega^2_{n-2}
\label{m1+}
\end{equation}
where $M(v)$ is the mass function and is related to the gravitational energy. This metric reduces to 
the Schwarzschild spacetime when $M(v)= M$, the mass of matter sphere as measured 
by an observer at infinity. 
From the point of view of the interior manifold $\it\bf(M^-)$, we can write down 
the timelike surface $\Sigma$ as
\begin{equation}
f_{-}(r,t)=r-r_{\Sigma}=0 \label{f1_}
\end{equation}
where $r_{\Sigma}$ is constant as $\Sigma$ is a comoving surface forming the boundary of the fluid.
Using Eq. (\ref{f1_}) in Eq. (\ref{m1}), we have the interior induced metric on $\Sigma$ is
\begin{equation}
ds^{2}_{-}=-d\tau^2+ R(r_{\Sigma},t)^2      d \Omega^2_{n-2}\,,\label{M_1}
\end{equation}
where $\tau$ is the proper time on $\Sigma$. Now, from the point of view of 
the exterior manifold $\it\bf(M^+)$, the boundary surface $\Sigma$ is
\begin{equation}
f_{+}(r,V)=r-r_{\Sigma}(v)=0. \label{f1+}
\end{equation}
Using Eq. (\ref{f1+}) in Eq. (\ref{m1+}), the induced metric on $\Sigma$ is
\begin{equation}
ds^{2}_{+}=-\left[1-\frac{2M(v)}{r^{2(N-1)}}+2\frac{dr_{_{\Sigma}}}{dv}\right]dv^2 + r^{2}\,. d \Omega^2_{n-2}
\label{M11+}
\end{equation}

Let us determine the extrinsic curvatures of $\Sigma$. The unit normal vectors to the $\Sigma$ from Eqs. (\ref{f1_}) and (\ref{f1+}) are given by
\begin{eqnarray}
n^{-}_{l}&=&\left[0,e^{\beta(r,t)},0,0\right]_{_\Sigma}, ~~~~~n^{+}_l=\left[1-\frac{2M\left(v\right)}{r^{2(N-1)}}+2\frac{dr}{dv}\right]^{-\frac{1}{2}}_\Sigma \left[-\frac{dr}{dv}\delta^0_l+\delta^1_l\right]_\Sigma .
\end{eqnarray}
The nonzero components of the extrinsic curvature on the either side of $\Sigma$ are
\begin{eqnarray}
K^{-}_{\tau\tau}&=&-\left[\alpha' e^{2\alpha(r,t)}e^{-\beta(r,t)}\left(\frac{dt}{d\tau}\right)^2\right]_{\Sigma} \label{td1}, ~~~~~K^{-}_{\theta\theta}=\left[R R' e^{-\beta}\right]_\Sigma \label{thd1}, ~~~~K^{-}_{\phi\phi}= \sin^2{\theta} K^{-}_{\theta\theta}\\
K^{+}_{\tau\tau}&=& \left[\frac{d^2v}{d\tau^2}\left(\frac{dv}{d\tau}\right)^{-1}-\left(\frac{dv}{d\tau}\right)\frac{M(v)}{r^{2N-1}}\right]_\Sigma \label{tu1}; ~
K^{+}_{\theta\theta}=\left[\left(\frac{dv}{d\tau}\right)\left\{1-\frac{2M(v)}{r^{2(N-1)}}\right\}r-r\frac{dr}{d\tau}\right]_\Sigma\label{thu1},\\
K^{+}_{\phi\phi}&=&\sin^2{\theta}\, K^{+}_{\theta\theta}.
\end{eqnarray}
For the junction conditions, we require
\begin{eqnarray}
ds^2_{\Sigma}&=&(ds_{-})^2_\Sigma = (ds_{+})^2_{\Sigma}, ~~
\left[K_{ij}\right]=K_{ij}^{+}-K_{ij}^{-}=0 .\label{MK}
\end{eqnarray}
For the metric coefficients, the matching leads to
\begin{eqnarray}
dt&=&e^{-\alpha(r,t)}d\tau \label{dt1}, ~~(r)_\Sigma=R(r_{_\Sigma},t) \label{Rr1}\\
\left(\frac{dv}{d\tau}\right)&=&\left[1-\frac{2M(v)}{r^{2(N-1)}}+2\frac{dr}{dv}\right]_{\Sigma}^{-1/2} \label{Vt1},
\end{eqnarray}
whereas, for Eqs. (\ref{thd1}) and (\ref{thu1}) in the junction condition Eq. \eqref{MK} on 
the hypersurface $\Sigma$, we get
\begin{equation}
	\left[R R' e^{-\beta}\right]_\Sigma=\left[\left(\frac{dv}{d\tau}\right)\left\{1-\frac{2M(v)}{r^{2(N-1)}}\right\}r-r\frac{dr}{d\tau}\right]_{\Sigma} . \label{kud1}
\end{equation}
Simplifying the above equation (\ref{kud1}) using Eqs. (\ref{Rr1}), (\ref{Vt1}) and the mass $F(r,t)$, we get
\begin{eqnarray}
2M(v)=F\left(r _{_\Sigma }\right)\label{F11}
\end{eqnarray}
For the matching with a Schwarzschild spacetime in the exterior, 
the above condition reduces to $2M=F(r)$
and has been used in the main text.
%%%%%%%%%%%%%%%%%%%%%%%%%%%%%%%%%%%%%%%%%%%%%%%%%%%%%%%%%%%%%%%%%%%%%%%%%%%%%%%%%%

\end{document}